\documentclass[epj,nopacs]{svjour}
\usepackage{amsmath,amssymb,amsfonts,graphicx,cite,multirow,color,subfig}
\usepackage[retainorgcmds]{IEEEtrantools}
\usepackage{ulem}
\usepackage{indentfirst}

\graphicspath{{graphics/}}
\makeatletter
\ifx\input@path\@undefined
\def\input@path{{graphics/}}
\else
\g@addto@macro\input@path{{graphics/}}
\fi
\makeatother

\newcommand{\sla}{\!\!\!\!\!\not\:\:\!}

\newcommand{\slac}{\!\!\!\!\!\!\!\!\!\!\not\,\,\,\,\,\,\,}

\newcommand{\HerwigPP}{\textsc{Herwig}\texttt{++}}
\newcommand{\Herwig}{\textsc{Herwig}}

\newcommand{\MG}{\textsc{MadGraph5\_aMC@NLO}}

  {\small
  \medmuskip=0mu
  \thinmuskip=0mu
  \thickmuskip=0mu
  \nulldelimiterspace=0pt
  \scriptspace=0pt
  \align}%
  {\endalign}

\preprint{CERN-TH-2018-154\\HERWIG-2018-02\\IPPP/18/55\\MCnet-18-12}

\title{Spin Correlations in Parton Shower Simulations}

\author{Peter Richardson\inst{1,2} and Stephen Webster\inst{2}}

\institute{
Theoretical Physics Department, CERN, 1211 Geneva 23, Switzerland
\and
IPPP, Department of Physics, Durham University}

\date{\today}

\abstract{Spin correlations are an important, but often neglected, effect in
  modern Monte Carlo event generators. We show that they can
  be fully incorporated in \Herwig7 using the algorithm originally proposed
  by Collins and Knowles in all stages of the event generation process
  and between the different stages of the event generation. In this paper
  we present the final missing ingredient, correlations
  in both the angular-ordered and dipole shower algorithms and between the
  parton shower and hard production and decay processes.}

\begin{document}

\authorrunning{P. Richardson and S. Webster}  \titlerunning{Spin Correlations}
\maketitle


\setlength{\parskip}{1ex}

\section{Introduction}

Monte Carlo event generators are essential in modern particle physics as they provide state-of-the-art theoretical calculations
which can be directly compared to experimental results\footnote{See Ref.\,\cite{Buckley:2011ms} for a recent review.}.
In recent years most of the developments in these programs have focused on improving the accuracy of the
hard perturbative calculation by matching with higher order and higher multiplicity matrix elements.
There has however been less work focused on improving the accuracy of the parton shower algorithm itself.
While there is some older work \cite{Kato:1986sg,Kato:1988ii,Kato:1990as,Kato:1991fs} there has recently been a
revival of interest in including higher order corrections in the parton shower evolution
in antenna \cite{Giele:2011cb,Hartgring:2013jma,Li:2016yez} and dipole \cite{Hoche:2017hno,Hoche:2017iem,Dulat:2018vuy} showers
as well as work on amplitude-based evolution to treat subleading colour effects \cite{Platzer:2012np,Martinez:2018ffw}.

One important, but often overlooked, effect is the azimuthal correlation of
emissions which occurs even at leading order in the parton shower evolution due to the polarization of the gluons.
Spin correlations also give rise to correlations between the emissions in the parton shower and the hard process, and
between the production and decay of heavy particles, such as the top quark. In order to fully include these
correlations, such that the complexity of the calculation only grows linearly with the number of
final-state particles, the algorithm of Refs.\,\cite{Collins:1987cp,Knowles:1987cu,Knowles:1988hu,Knowles:1988vs,Richardson:2001df}
can be used. This algorithm was used in \Herwig6~\cite{Corcella:2000bw} to implement the correlations in the parton shower, 
between the hard process and the parton shower emissions~\cite{Knowles:1987cu,Knowles:1988hu,Knowles:1988vs} and between the
production and decay of heavy particles, in both the Standard Model, the MSSM and its R-parity violating
extension.\footnote{However correlations between the parton shower and decays were not included for technical reasons.}
The same algorithm is also used in the EvtGen package\cite{Lange:2001uf} for correlations in the decays of hadrons.

In \HerwigPP\cite{Bahr:2008pv} and \Herwig7 \cite{Bellm:2015jjp,Bellm:2017bvx} this algorithm was used to provide
correlations between the production and decay of heavy particles, and in the decay of hadronic resonances, in a unified
framework. However the correlations in the parton shower were not included. In this paper we will present the necessary
calculations in order to implement these correlations in both parton shower algorithms available in \Herwig7
and results from their implementation which will be available in  \Herwig7.2.

In the next section we will first recap the details of the algorithm of
Refs.\,\cite{Collins:1987cp,Knowles:1987cu,Knowles:1988hu,Knowles:1988vs,Richardson:2001df} and then present details of
the calculations needed for its implementation in the \Herwig7 parton showers. We then present some results
from the implementation in \Herwig7 followed by our conclusions.

\section{Algorithm}
\label{sect:algorithm}

\subsection{Review of the Algorithm}
\label{sect:algorithm-review}

It is best to consider the details of the algorithm of
Refs.\,\cite{Collins:1987cp,Knowles:1987cu,Knowles:1988hu,Knowles:1988vs,Richardson:2001df} by considering an example.
In order to consider the correlations in the parton shower we will consider the example of the decay of the Higgs
boson to two gluons, followed by the branching of the gluons into quark-antiquark pairs, {\it i.e.}
$h^0\to gg\to q\bar{q}q'\bar{q'}$.

The algorithm proceeds by first generating the momenta of the gluons produced in the Higgs boson decay using the
appropriate matrix element. Obviously in this case this is trivial, however the structure of
the matrix element will give rise to the subsequent correlations. One of the outgoing gluons is then selected at random and
a spin density matrix calculated for its subsequent evolution
\begin{equation}
  \rho_{g_1}^{\lambda_{g_1}\lambda_{g_1}'} = \frac1N\mathcal{M}^{\lambda_{g_1}\lambda_{g_2}}_{h^0\to gg}{\mathcal{M}^*}^{\lambda'_{g_1}\lambda_{g_2}}_{h^0\to gg},
\end{equation}  
where $\mathcal{M}_{h^0\to gg}$ is the helicity amplitude for the decay of the Higgs boson into two gluons,
$\lambda_{g_1}$ and $\lambda'_{g_1}$ are the polarizations of the first gluon and $\lambda_{g_2}$ is the polarization
of the second gluon. The Einstein convention where repeated indices are summed over is used.
The normalization $N$ is defined such that the trace of the spin density matrix \mbox{${\rm Tr}\,\rho_{g_1}=1$} and is therefore
equal to the spin-averaged matrix element squared used to calculate the momenta of the gluons in the previous
stage of the algorithm.

We can now proceed and generate the parton shower emissions from the gluon in the standard way because the
probability of such an emission occurring is not affected by the spin density
matrix\footnote{While this is true for emissions in both QCD and QED if we were to consider the emission of electroweak $W^\pm$ and $Z^0$
  bosons then the emission probability will also depend on the spin density matrix as the diagonal elements give the probability of having a left- or right-handed particle radiating.}. However the azimuthal angle of the emission is affected by the spin density matrix and is generated using the distribution
\begin{equation}
  \rho_{g_1}^{\lambda_{g_1}\lambda_{g_1}'} \mathcal{M}^{\lambda_{g_1}\lambda_q\lambda_{\bar{q}}}_{g\to q\bar{q}}\mathcal{M}^{*\lambda'_{g_1}\lambda_q\lambda_{\bar{q}}}_{g\to q\bar{q}},
  \label{eq:split1}
\end{equation}
where $\mathcal{M}_{g\to q\bar{q}}$ is the helicity amplitude for the splitting of a gluon into a quark-antiquark pair
and $\lambda_q$ and $\lambda_{\bar{q}}$ are the helicities of the quark and antiquark, respectively.
Here we have used the spin-density matrix to encode the correlations from the hard process, rather than just averaging over the
polarizations of the branching gluon.

This relies on the standard factorization of the matrix element in the collinear limit
into the production of an on-shell gluon and its subsequent branching, which forms the
basis of the parton shower approach. Similarly when generating the decay of a particle the
amplitude factorizes in the narrow width limit into an amplitude for the production
of the particle and one for its decay.

In the case that we have further emissions from the quarks produced in the gluon branching we would then calculate a spin density matrix
for emission from the\linebreak quarks in the same way as above, {\it i.e.}
\begin{equation}
  \rho_q^{\lambda_q\lambda'_q} = \frac1N\rho_{g_1}^{\lambda_{g_1}\lambda_{g_1}'} \mathcal{M}^{\lambda_{g_1}\lambda_q\lambda_{\bar{q}}}_{g\to q\bar{q}}\mathcal{M}^{*\lambda'_{g_1}\lambda'_q\lambda_{\bar{q}}}_{g\to q\bar{q}},
\end{equation}
where again the normalization is such that \mbox{${\rm Tr}\,\rho_q=1$} and is equal to the spin contracted amplitude used to
calculate the distribution of the $g\to q\bar{q}$ splitting, Eqn.\,\ref{eq:split1}.

Once there is no further emission, or decay of the particle, we can compute a decay matrix for the final parton shower branching
or decay. In our example if there is no further emission from the quark or antiquark then the decay matrix for the gluon is
\begin{equation}
  D^{\lambda_{g_1}\lambda'_{g_1}}_{g_1} = \frac1N \mathcal{M}^{\lambda_{g_1}\lambda_q\lambda_{\bar{q}}}_{g\to q\bar{q}}\mathcal{M}^{*\lambda'_{g_1}\lambda_q\lambda_{\bar{q}}}_{g\to q\bar{q}}, 
\end{equation}
where again the normalization is such that \mbox{${\rm Tr} D_{g_{1}}=1$}. If there had been emissions from either the quark or antiquark
then instead of summing over their helicities we would contract with the appropriate decay matrix. These decay matrices
encode the distributions generated in the branchings so that any subsequent emissions are correctly correlated with them.

We now need to generate the branching of the second gluon from the hard process. We first calculate the
spin density matrix, however now instead of summing over the helicities of the other gluon we contract with
the decay matrix encoding how the first gluon branched, {\it i.e.}
\begin{equation}
  \rho_{g_2}^{\lambda_{g_2}\lambda'_{g_2}} = \frac1N\mathcal{M}^{\lambda_{g_1}\lambda_{g_2}}_{h^0\to gg}{\mathcal{M}^*}^{\lambda'_{g_1}\lambda'_{g_2}}_{h^0\to gg}
       D_{g_1}^{\lambda_{g_1}\lambda'_{g_1}},
\end{equation}
where again the normalization is such that \mbox{${\rm Tr}\rho_{g_2}=1$}.

As before this spin density matrix is then contracted with the matrix elements for the branching of the gluon to
determine the azimuthal angle according to Eqn.\,\ref{eq:split1} with $g_1\to g_2$.

It is worth noting that due to the choice of the normalization of the spin density matrices the normalization used in the
next step of the calculation is always equal to the distribution used to generate the previous step. This ensures that the final
distribution is equal to the full result, up to the approximation used to factorize the full matrix element into different components.
In the example we have been considering the final result used to generate the momenta of the particles is
\begin{equation}
                     \mathcal{M}^{\lambda_{g_1}\lambda_{g_2}}_{h^0\to gg}{\mathcal{M}^*}^{\lambda'_{g_1}\lambda'_{g_2}}_{h^0\to gg} \mathcal{M}^{\lambda_{g_1}}_{g\to q\bar{q}}\mathcal{M}^{*\lambda'_{g_1}}_{g\to q\bar{q}} \mathcal{M}^{\lambda_{g_2}}_{g\to q'\bar{q'}}\mathcal{M}^{*\lambda'_{g_2}}_{g\to q'\bar{q'}},
\end{equation}
    {\it i.e.} the full matrix element for the process in the collinear limit.

    The same arguments also apply for initial-state radiation with the r\^{o}les of the spin
    density and decay matrices exchanged, as we evolve backwards from the hard
    process~\cite{Collins:1987cp,Knowles:1987cu,Knowles:1988hu,Knowles:1988vs,Richardson:2001df}.
    A similar example, considering the production and decay of a top quark-antiquark pair can be found in
    Ref.\,\cite{Bahr:2008pv}.

    It is worth noting that we could consider the spin sum used where a
    branching or decay has not been generated as contracting with a spin density or decay
    matrix proportional to the identity matrix. This can be generalised, {\it i.e.} if we have
    information on the production, decay or branching of a particle we should use the appropriate
    spin density or decay matrix.
    This allows us to change the order in which we calculate the decay or branchings if required. For
    example, we postpone the generation of any tau lepton decays until after the parton shower has been
    generated to avoid applying large boosts to the low invariant mass systems produced which leads
    to numerical issues with momentum conservation. This has the price of increasing the number
    of numerical evaluations as any spin density and decay matrices which are affected by the
    decay then need to be recalculated to ensure consistency, however this is still a linear operation.
    This is also required in the dipole shower where successive emissions can be from different particles.

\subsection{Helicity Amplitudes for Parton Branching}
\label{sect:branching}
In order to implement the algorithm described above we need the helicity amplitudes for the branching
processes in the same limit as that used in the parton shower. In modern parton shower algorithms
this is the {\it quasi}-collinear limit of Ref.\,\cite{Catani:2000ef}. In order to
efficiently implement the algorithm we need compact analytic expressions for these
branchings, with the correct approximations,\linebreak rather than relying on numerical evaluations of the
helicity amplitudes, which are used in \Herwig7 for the amplitudes for production and decay processes.

We choose to calculate the branchings in a frame where the branching particle is along the $z$-axis,
which is the choice made in the angular-ordered parton shower algorithm~\cite{Gieseke:2003rz} in \Herwig7.
In this frame the momentum of the branching particle is
\begin{equation}
  p =   (\sqrt{{\bf p}^2+m_0^2}; 0,0,{\bf p}),
\end{equation}
where $m_0$ is the on-shell mass of the branching particle and ${\bf p}$ is the magnitude of its three-momentum.
In the parametrization used in \Herwig7 all momenta are decomposed in the Sudakov basis,
{\it i.e.} the momentum of particle $i$ can be written as
\begin{equation}
  q_i = \alpha_ip+\beta_in +q_{\perp i},
  \label{eqn:herwigmomentum}
\end{equation}
where $p$ is the momentum of the branching particle before any emission, $n$ is a light-like reference
vector, $\alpha_i$ and $\beta_i$ are coefficients, and $q_{\perp i}$ is the transverse momentum relative to the directions
of $p$ and $n$. The reference vectors and transverse momenta satisfy $p^2=m^2_0$, $n^2=0$, $p\cdot q_{\perp i}=n\cdot q_{\perp i}=0$
and $p\cdot n>0$.
 As we need an explicit representation we take
\begin{align}
 n &= (1,0,0,-1).
\end{align}

If we consider a branching $0\to1,2$ then in
this parameterization the momenta of the particles produced in the branching are:
\begin{subequations}
\begin{eqnarray}
  q_1 &=& zp+\beta_1n +q_\perp; \\
  q_2 &=& (1-z)p+\beta_2n -q_\perp;
\end{eqnarray}
where $z$ is the light-cone momentum fraction of the first parton and
\begin{eqnarray}
  q_\perp &=& (0;p_\perp\cos\phi,p_\perp\sin\phi,0); \\
  \beta_1 &=& \frac1{2zp\cdot n}\left(p_\perp^2+m_1^2-z^2m_0^2\right); \\
  \beta_2 &=& \frac1{2(1-z)p\cdot n}\left(p_\perp^2+m_2^2-(1-z)^2m_0^2\right);
\end{eqnarray}
where $m_{1,2}$ are the on-shell masses of the particles produced, $p_\perp$ is the magnitude of
the transverse momentum of the branching and $\phi$ is the azimuthal angle for the first particle
produced in the branching.

In the \Herwig7 angular-ordered parton shower we use the evolution variable
\begin{equation}
  \tilde{q}^2 = \frac{q^2_0-m^2_0}{z(1-z)},
\end{equation}
for final-state evolution.

The momentum of the off-shell particle initiating the branching is
\begin{equation}
  q_0 = p +\beta_0n,
\end{equation}
where
\begin{equation}
  \beta_0 = \beta_1+\beta_2 = \frac1{2p\cdot n}\left(\frac{p_\perp^2}{z(1-z)}+\frac{m_1^2}z+\frac{m_2^2}{1-z}-m_0^2\right),
\end{equation}
such that the virtuality of the branching parton is
\begin{equation}
  q_0^2 = \frac{p_\perp^2}{z(1-z)}+\frac{m_1^2}z+\frac{m_2^2}{1-z}.
\end{equation}
\end{subequations}
We need to evaluate the branchings in the {\it quasi}-collinear limit in which we take the masses and
transverse momentum to zero while keeping the ratio of the masses to the transverse momentum fixed.
Practically this is most easily achieved by rescaling the masses and transverse momentum by a parameter
$\lambda$ and expanding in $\lambda$ \cite{Catani:2000ef}, {\it i.e.}
\begin{align}
  m_i\to&\lambda m_i & \& &    &p_\perp\to&\lambda p_\perp.
\end{align}

In order to define our conventions for spinors and polarization vectors we present the calculation
of the $g\to q\bar{q}$ splitting here. The results for the remaining QCD branchings are presented in Appendix
\,\ref{app:split}.

In this case we start with an on-shell gluon such that after
rotating so that the gluon is along the $z$-direction
the momentum and polarization vectors of the gluon are:
\begin{eqnarray}
 p                        &=& {\bf p}(1; 0, 0, 1);\\
 \epsilon^\mu_{\lambda_0} &=& \frac1{\sqrt{2}}(0; -\lambda_0, -i, 0);
\end{eqnarray}
where $\lambda_0=\pm1$ is the helicity of the gluon.
For the branching $g\to q \bar{q}$ , $m_1=m_2=m$, therefore
expanding in $\lambda$ the momenta of the particles in the branching are:
\begin{subequations}
\begin{eqnarray}
  q_0 &=& \left( {\bf p}+\frac{m^2+{p_\perp}^2}{4{\bf p}z\left(1-z\right)}\lambda^2  , 0,0, {\bf p}-\frac{m^2+p_\perp^2}{4{\bf p}z\left(1-z\right)}\lambda^2\right)\nonumber\\&& +\mathcal{O}(\lambda^3);\\
  q_1 &=& \left( z{\bf p}+ \frac{m^2+p_\perp^2}{4z{\bf p}}\lambda^2, \lambda p_\perp\cos\phi, \lambda p_\perp\sin\phi,\right.\nonumber \\&&\left.  z{\bf p}-\frac{m^2+p_\perp^2}{4z{\bf p}}\lambda^2\right)+\mathcal{O}(\lambda^3);\\
  q_2 &=& \left( \left(1-z\right){\bf p}+\frac{m^2+p_\perp^2}{4\left(1-z\right){\bf p}}\lambda^2,-\lambda p_\perp\cos\phi, -\lambda  p_\perp\sin\phi,\right.\nonumber\\
  && \ \ \  \ \ \left.
  \left(1-z\right){\bf p}-\frac{m^2+p_\perp^2}{4\left(1-z\right){\bf p}}\lambda^2 \right)
  +\mathcal{O}(\lambda^3).
\end{eqnarray}
\end{subequations}

 In general we define the spinors using the conventions in \cite{Haber:1994pe}.
 This allows us to write the spinors for the outgoing particles:
 \begin{subequations}
\begin{eqnarray}
  \bar{u}_{\frac12}(q_1)   &=& \left[
    \sqrt {2{\bf p}z},
    \frac {{\rm e}^{-i\phi}p_\perp}{\sqrt {2{\bf p}z}}\lambda,
    \frac{m}{\sqrt{2{\bf p}z}}\lambda,
\frac { p_\perp m\lambda^2 {\rm e}^{-i\phi}}{\left[2z{\bf p}\right]^{3/2}}
    \right];\nonumber\\
  \bar{u}_{-\frac12}(q_1) & =& \left[
    -\frac {p_\perp m \lambda^2{\rm e}^{i\phi}}{\left[2z{\bf p}\right]^{3/2}},
    \frac{m}{\sqrt {2{\bf p}z}}\lambda,
    -\frac{{\rm e}^{i\phi}p_\perp}{\sqrt{2{\bf p}z}}\lambda,
    \sqrt {2{\bf p}z}
    \right];\nonumber\\&&
\end{eqnarray}
and
\begin{equation}
  v_{\frac12}(q_2) = \left[\begin{array}{c}-\frac {p_\perp {\rm e}^{-i\phi} }{\sqrt {2{\bf p}(1-z)}}\lambda\\
      -\sqrt {2{\bf p}(1-z)}\\
      {\frac {mp_\perp\lambda^2 {\rm e}^{-i\phi} }{\left[(2{\bf p}(1-z)\right]^{3/2}}}\\
      \frac {m}{\sqrt {2{\bf p}(1-z)}}\lambda
    \end{array}\right]; \ \
   v_{-\frac12}(q_2) = \left[\begin{array}{c}\frac{m}{\sqrt{2{\bf p}(1-z)}}\lambda\\
       -\frac {mp_\perp\lambda^2 {\rm e}^{i\phi} }{\left[2{\bf p}(1-z)\right]^{3/2}}\\
       -\sqrt {2{\bf p}(1-z)}\\
       \frac {p_\perp {\rm e}^{i\phi} }{\sqrt {2{\bf p}(1-z)}}\lambda\end{array}\right].
\end{equation}
\end{subequations}
 
 We define the helicity amplitudes $F_{\lambda_0\lambda_1\lambda_2}$ for the branching 
 such that the spin averaged splitting function is
\begin{equation}
 P(z) = \frac12\sum_{\lambda_0,\lambda_1,\lambda_2} F_{\lambda_0\lambda_1\lambda_2}
					     F^*_{\lambda_0\lambda_1\lambda_2}.
\end{equation}
For the $g\to q\bar{q}$ branching
\begin{equation}
F_{\lambda_0\lambda_1\lambda_2} =  \sqrt{\frac{q_0^2}{2}}\bar{u}_{\lambda_1}(q_1) \epsilon_{\lambda_0}\slac v_{\lambda_2}(q_2).
\end{equation}
There is no simple form for all the helicities and therefore the functions, 
$F_{\lambda_0\lambda_1\lambda_2}$, are given in Table~\ref{tab:gluon}.
These amplitudes give the correct spin averaged massive splitting 
function and reduce to the splitting functions used in \Herwig6 from \cite{Knowles:1987cu}~Table~1
in the massless limit.\footnote{N.B. the $g\to g g$ and $g\to q \bar{q}$ branchings require a phase
  change to give complete agreement with~\cite{Knowles:1987cu} but this phase is arbitrary and does
  not affect any physical results.}

\subsection{Examples}
\label{sect:examples}
  
It is instructive to calculate the correlations in
some simple cases. This will both illustrate how the spin correlation
algorithm works and provide analytical results that we later use to
check the general implementation in \Herwig7.

The simplest example is the correlation of the angle between the planes of
two successive parton shower branchings, {\it i.e.} $0\to12$ followed by $2\to34$.
The only non-zero correlation is due to the polarization of an intermediate gluon.

If we consider the branching $q\to qg$ the spin density matrix for the radiated gluon is
\begin{equation}
  \rho_g = \left(\begin{array}{cc}\
    \frac12
    & -\frac {z_1{\rm e}^{2i\phi_1}}{1+z_1^2}\\
    -\frac {z_1{\rm e}^{-2i\phi_1}}{1+z_1^2}
    & \frac12\end{array}\right),
\end{equation}
where $z_1$ is the momentum fraction and $\phi_1$ the azimuthal angle of the quark, respectively.
We have neglected the mass of the radiating quark.

Similarly for the branching $g\to gg$ the spin density matrix for the radiated gluon is
\begin{equation}
  \rho_g = \left(\begin{array}{cc}\
    \frac12
    &-\frac {z_1^2{\rm e}^{ 2i\phi_1}}{2\left(1-z_1(1-z_1)\right) ^2}\\
     -\frac {z_1^2{\rm e}^{-2i\phi_1}}{2\left(1-z_1(1-z_1)\right) ^2}    
    & \frac12\end{array}\right).
\end{equation}

\begin{table}
  \begin{center}
  \begin{tabular}{|c|c|c|c|}
    \hline
    First & Second & $A$ & $B$ \\
    Branching & Branching & & \\
    \hline
    $q\to qg$ & $g\to q\bar{q}$ & $\frac{2z_1}{1+z_1^2}$ & $\frac{-2z_2(1-z_2)}{1-2z_2(1-z_2)}$  \\
    \hline
    $q\to qg$ & $g\to gg$       & $\frac{2z_1}{1+z_1^2}$ & $\frac{(z_2(1-z_2))^2}{(1-z_2(1-z_2))^2}$ \\
    \hline
    $g\to gg$ & $g\to q\bar{q}$ & $\frac{z_1^2}{(1-z_1(1-z_1))^2}$ & $\frac{-2z_2(1-z_2)}{1-2z_2(1-z_2)}$\\
    \hline
    $g\to gg$ & $g\to gg$       & $\frac{z_1^2}{(1-z_1(1-z_1))^2}$ & $\frac{(z_2(1-z_2))^2}{(1-z_2(1-z_2))^2}$ \\
    \hline
  \end{tabular}
  \caption{The coefficients $A$ and $B$ for the correlation between the azimuthal angles of the first and second branching. The momentum
           fraction in the first branching is $z_1$ and $z_2$ in the second branching.}
  \label{tab:nearest}
  \end{center}
\end{table}
If we contract these spin density matrices with the appropriate helicity amplitudes
for the subsequent branching of the gluon then we obtain the distribution
\begin{equation}
  \frac1{2\pi}\left[1+A\,B\cos2\Delta\phi\right],
  \label{eq:nearest}
\end{equation}
where $\Delta\phi=\phi_2-\phi_1$ is the angle between the planes of the two branchings and the coefficients
$A$ and $B$ are given in Table\,\ref{tab:nearest}.

The simplest example in which there are final-state correlations
in the parton shower due to the production process
is the decay of the Higgs boson to two gluons followed by the
branching of each of the two gluons into a quark-antiquark pair, {\it i.e.}
$h^0\to g g \to q \bar{q} q'\bar{q'}$, which we have already considered.

We take the amplitude to be\footnote{This is the correct form in the infinite top-mass limit and
any changes from including the finite top mass would cancel in the normalized distribution.}
\begin{equation}
  \mathcal{M}_{h^0\to gg} = -p_1\cdot p_2 \epsilon^*_1\cdot \epsilon_2^* +p_2\cdot\epsilon_1^* p_1\cdot\epsilon_2^*,
\end{equation}
where $p_{i=1,2}$ and $\epsilon_{i=1,2}$ are the 4-momenta and polarization vectors of the outgoing
gluons, respectively. We have neglected the overall normalization of the matrix element
which affects the total rate but not the correlations we are studying.

The non-zero helicity amplitudes for $h^0\to g g$ are:
\begin{subequations}
\begin{eqnarray}
  \mathcal{M}_{h^0\to gg}^{++} &=& -\frac{m_h^2}2{\rm e}^{-2i\phi};\\
  \mathcal{M}_{h^0\to gg}^{--} &=& -\frac{m_h^2}2{\rm e}^{ 2i\phi};
\end{eqnarray}
\end{subequations}
where $\phi$ is the azimuthal angle of the 1st gluon.

We can use the helicity amplitudes in Table\,\ref{tab:gluon} to calculate the decay matrix
for the first gluon branching
\begin{equation}
  D = \left(\begin{array}{cc} \frac12 & a_1(z_1,\tilde{q}_1){{\rm e}^{2\,i\phi_{1}}}\\
a_1(z_1,\tilde{q}_1){{\rm e}^{-2\,i\phi_{1}}} &  \frac12  \end{array}\right),
\end{equation}
where $\phi_1$ is the azimuthal angle of the quark $q$ in a frame in which the 1st gluon is along the $z$-axis and
\begin{equation}
  a_1(z_1,\tilde{q}_1)=
    \frac{ \left( z_1 \left( 1-z_1 \right) -{\frac {m_q^2}{z_1 \left( 1-z_1 \right) {\tilde{q}_1}^2}} \right)}
         {\left( 1- 2z_1 \left( 1-z_1 \right) +\frac {2m_q^2}{z_1 \left( 1-z_1 \right) {\tilde{q}_1}^2} \right)},
\end{equation}
where $z_1$ is the energy fraction of the quark, $\tilde{q}_1$
the evolution variable for the branching and $m_q$ is the mass of the
quark.

We can then calculate the spin density matrix
needed to calculate the second branching,
\begin{equation}
  \rho =\left(\begin{array}{cc} \frac12 & a_1(z_1,\tilde{q}_1){{\rm e}^{2i\phi_{1}-4i\phi}}\\
       a_1(z_1,\tilde{q}_1){{\rm e}^{-2i\phi_{1}+4i\phi}} &  \frac12  \end{array}\right).
\end{equation}
This gives the distribution
\begin{eqnarray}
  \left( 1-2z_2 \left( 1-z_2 \right) +\frac {2m^2_{q'}}{z_2 \left( 1-z_2 \right) {\tilde{q}_2}^2} \right)\nonumber \\
  \left( 1+4a_1(z_1,\tilde{q}_2)a_{2}(z_2,\tilde{q}_2)\cos \left(4\phi  -2\phi_2-2\phi_1\right)  \right), 
\end{eqnarray}
where $z_2$ is the energy fraction of the quark $q'$ and $\tilde{q}_2$ the evolution variable for the second branching.
The azimuthal angle of the second branching, $\phi_2$, is measured in a frame where the 2nd gluon is along the
$z$-direction. If we rotate the quark produced in the first branching into this frame its angle $\phi'_1=2\phi-\phi_1+\pi$.

If we neglect the mass of the quark and multiply the distribution by the spin-averaged splitting
function for the first branching we obtain the distribution
\begin{equation}
  P_{g\to q\bar{q}}(z_1)P_{g\to q\bar{q}}(z_2) \left[ 1+4a_1(z_1)a_{2}(z_2)\cos\left(2\left[\phi_2-\phi'_1\right]\right)  \right].
\end{equation}
Integrating over $z_{1,2}$ between $0$ and $1$ and normalizing the resulting distribution gives
\begin{equation}
  \frac1{8\pi} \left[3+2\cos^2\left( \Delta\phi\right)\right],
  \label{eq:h0-gg-qqqq}
\end{equation}
where $\Delta\phi=\phi_2-\phi'_1$ is the azimuthal angle between the planes of the two branchings. 

\subsection{Implementation in the Angular-Ordered Parton Shower}
\label{sect:AO-shower}

The algorithm described in Section\,\ref{sect:algorithm-review} is ideal for implementation
in the angular-ordered parton shower in \Herwig7. In the angular-ordered parton shower each parton
from a hard process or decay is showered independently and the individual splittings are calculated in
the same order as that described in Section~\ref{sect:algorithm-review}.

There is one complication.
In the angular-ordered\linebreak shower,
each parton incoming to or outgoing from the hard process
is showered in a specific frame. The transformation between the production frame and showering frame
of a parton can introduce a rotation of the basis states of the parton which must be taken into
account in the treatment of the spin density and decay matrices of the parton.
This is done through the application of `mappings' to the matrices and is described in detail in
Appendix\,\ref{sect:rotation}.

\subsection{Implementation in the Dipole Shower}
\label{sect:dipole-shower}

Several additional considerations are necessary to implement the spin correlation
algorithm in the dipole shower.

A dipole consists of two colour-connected external partons, an emitter and a spectator.
In general the spectator defines the soft radiation pattern and, for several types of dipole,
is used to absorb the recoil momentum in a splitting. In other cases a set of outgoing
particles is used to absorb the recoil momentum in splittings.
The algorithm described in this paper does not include any formal treatment for spectators
or the recoil in splittings.
In each splitting the momentum of one or several particles are therefore modified in some way
that is not described by the helicity amplitudes.
One consideration that we do make is to ensure that any change in the basis states of a spectator,
or other particle, arising from the change in its momentum is accounted for.
In Section\,\ref{sect:results} we investigate the effect of the treatment of
recoils on predictions made using the dipole shower.

The objects that evolve in the dipole shower are series of colour-connected dipoles, referred to
as `dipole chains' \cite{Platzer:2011bc}. At a given point in the shower evolution
each dipole in the selected chain has a probability to radiate.
This is in contrast to the treatment in the angular-ordered shower in which the shower evolution
proceeds independently for each parton from a hard process or decay.
Extra care must therefore be taken in the dipole shower to ensure that the spin density
and decay matrices required for the computation of a given splitting are up-to-date.
Furthermore, in the dipole shower, each splitting takes place in a unique frame defined by the momenta
of the emitter and spectator. We must therefore compute and apply the mappings described in
Appendix\,\ref{sect:rotation} for every splitting in the dipole shower.

The splitting probabilities in the dipole shower are given by the
Catani-Seymour splitting kernels \cite{Catani:1996vz,Catani:2002hc}. These splitting kernels,
and the kinematics used to describe splittings in the dipole shower, are written in terms of
splitting variables specific to this formulation.
On the other hand the helicity amplitudes in Section\,\ref{sect:branching} and
Appendix\,\ref{app:split} are derived using a decomposition of the splitting momenta into the
Sudakov basis. In particular the helicity amplitudes are written in terms of the light-cone
momentum fraction $z$.
In practice in the dipole shower in \Herwig7 we actually generate the variables $z$ and $p_\perp$,
the transverse momentum of the emission, and map these to the variables required to express the
splitting kernels. No additional work is therefore required to use the helicity amplitudes in
Table\,\ref{tab:gluon} and Table\,\ref{tab:quark} in the dipole shower.

\section{Results}
\label{sect:results}

\subsection{Correlations inside the Parton Shower}
\label{sect:corr-in-shower}
\begin{figure} 
  \begin{center}
    \includegraphics[width=0.45\textwidth]{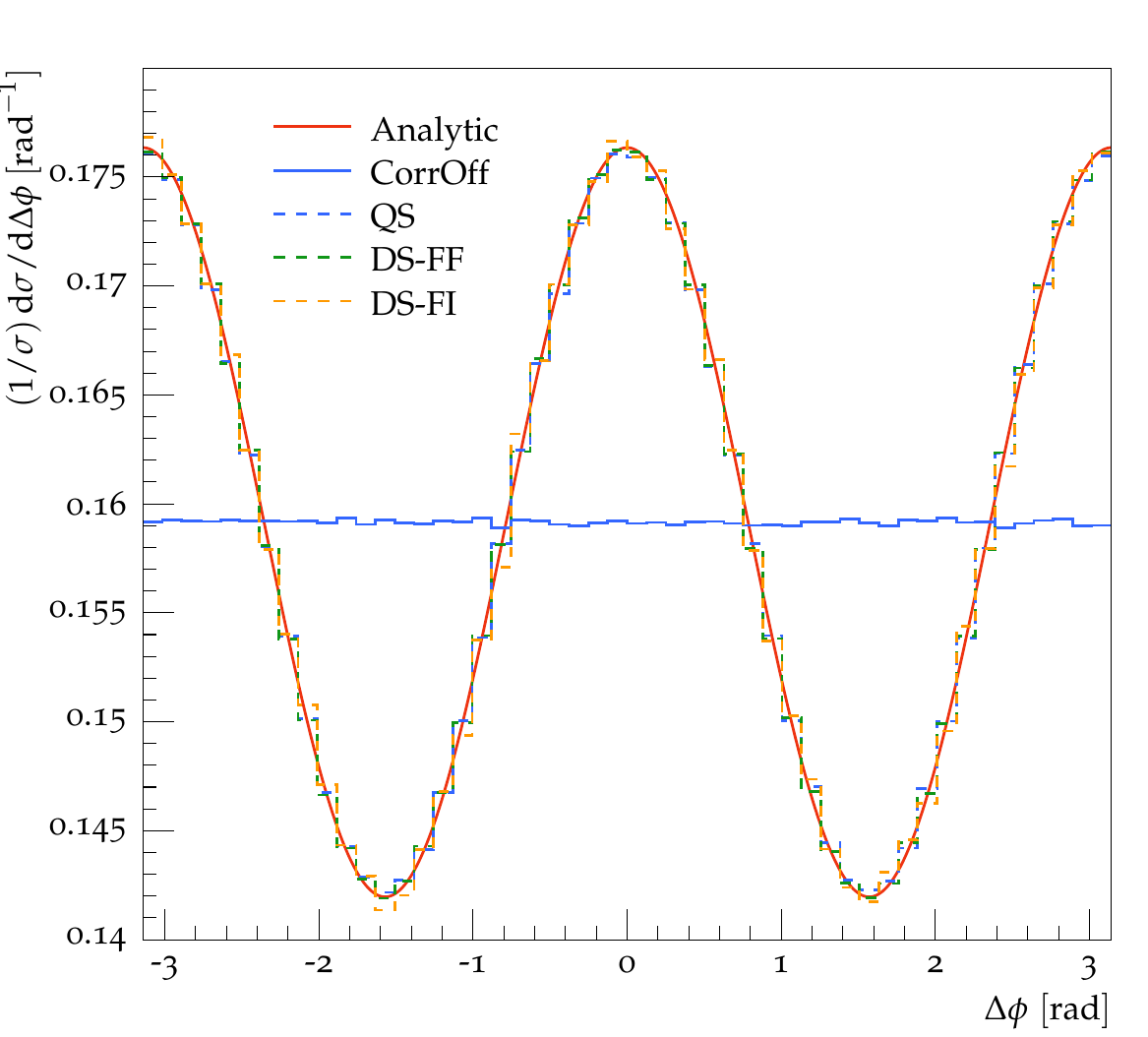}
  \end{center}
  \vspace{-10pt}
  \caption{The analytic result for the difference in azimuthal angle between
    the branching planes of subsequent final-state $q\to qg$ and $g\to gg$ splittings
    compared to the distributions predicted using the angular-ordered~(QS) and
    dipole parton showers in \Herwig7. The predictions obtained using only final-final~(DS-FF)
    and final-initial~(DS-FI) dipoles in the dipole shower are shown separately.
    The prediction of the angular-ordered~(CorrOff) shower without spin correlations is included
    for comparison.
    The momentum fraction in the first and second branchings lies in the range
    $0.9 < z_1 < 1.0$ and $0.4 < z_2 < 0.5$ respectively.}
  \label{fig:FSR-NearestNeighbour-qg-gg}
\end{figure}

The analytic result for the distribution of the angular difference between the planes of successive
parton shower branchings is given in Eqn.~\eqref{eq:nearest}. This expression can be expanded for
each of the four sequences of branchings that give rise to correlations using the expressions in
Table\,\ref{tab:nearest}. In this section we present the predictions
of these distributions obtained using the angular-ordered and dipole parton showers in \Herwig7.
The angular difference between two successive parton shower branchings is measured in the splitting
frame of the second branching.
This test verifies the implementation of the helicity amplitudes in both parton showers.
Furthermore, in the dipole shower, it also probes the implementation of the basis state mappings
between splittings.

\begin{figure}
  \begin{center}
    \includegraphics[width=0.45\textwidth]{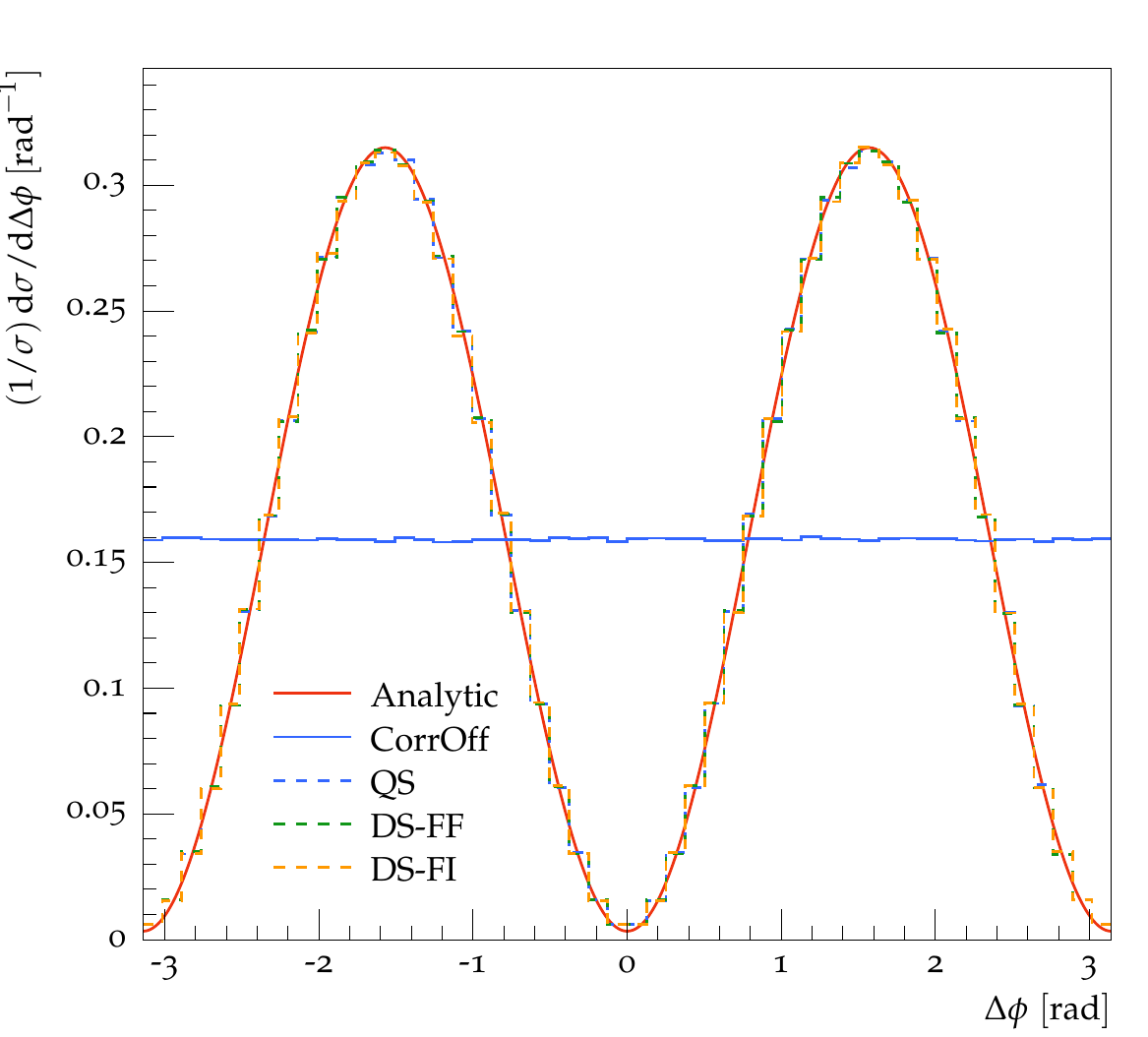}
  \end{center}
  \vspace{-10pt}
  \caption{The analytic result for the difference in azimuthal angle between
    the branching planes of subsequent final-state $q\to qg$ and $g\to q\bar{q}$ splittings
    compared to the distributions predicted using the angular-ordered~(QS) and
    dipole parton showers in \Herwig7. The predictions obtained using only final-final~(DS-FF)
    and final-initial~(DS-FI) dipoles in the dipole shower are shown separately.
    The prediction of the angular-ordered~(CorrOff) shower without spin correlations is included
    for comparison.
    The momentum fraction in the first and second branchings lies in the range
    $0.9 < z_1 < 1.0$ and $0.4 < z_2 < 0.5$ respectively.}
  \label{fig:FSR-NearestNeighbour-qg-qq}
\end{figure}
We test the cases of final-state radiation (FSR) and initial-state radiation (ISR) separately. 
In the ISR case the first splitting is identified as that closest to the incoming hadron
and the intermediate parton between the splittings is space-like.
In the dipole shower we can divide FSR and ISR further according to the type of dipole considered.
Specifically we divide FSR, radiation from a final-state emitter, according to whether the
spectator is final-state (final-final) or initial-state (final-initial) and we divide ISR,
radiation from an initial-state emitter, according to whether the spectator is final-state
(initial-final) or initial-state (initial-initial).
We include a separate result for each of these four types of dipole.
Note that we do not consider radiation from decay processes in this test.

For the purpose of these tests we implement an artificial restriction on the splittings allowed
in the dipole shower. Following the first splitting we only allow subsequent splittings from dipoles
in which the spectator is the spectator of the previous splitting and in which the emitter was
produced as a new parton in the previous splitting.
This restriction has two purposes. First, by forbidding subsequent emissions from different
legs of the hard process, it allows us to probe only the correlations in the shower. Second,
by using the same spectator in subsequent splittings the frame of the second splitting is
a suitable frame in which to measure the angular difference between the splittings.

\begin{figure}
  \begin{center}
    \includegraphics[width=0.45\textwidth]{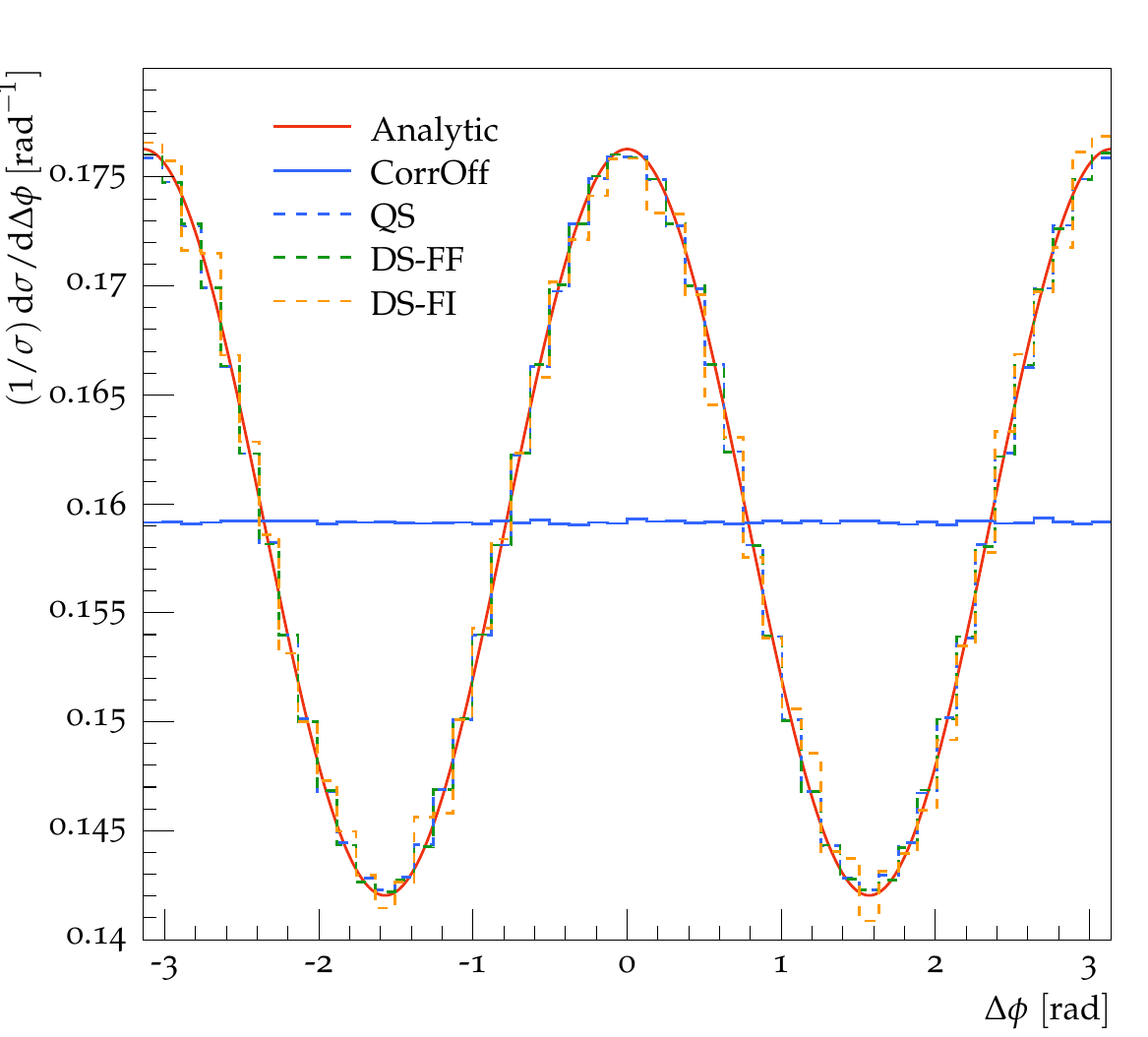}
  \end{center}
  \vspace{-10pt}
  \caption{The analytic result for the difference in azimuthal angle between
    the branching planes of subsequent final-state $g\to gg$ and $g\to gg$ splittings
    compared to the distributions predicted using the angular-ordered~(QS) and
    dipole parton showers in \Herwig7. The predictions obtained using only final-final~(DS-FF)
    and final-initial~(DS-FI) dipoles in the dipole shower are shown separately.
    The prediction of the angular-ordered~(CorrOff) shower without spin correlations is included
    for comparison.
    The momentum fraction in the first and second branchings lies in the range
    $0.9 < z_1 < 1.0$ and $0.4 < z_2 < 0.5$ respectively.}
  \label{fig:FSR-NearestNeighbour-gg-gg}
\end{figure}

\begin{figure}
  \begin{center}
    \includegraphics[width=0.45\textwidth]{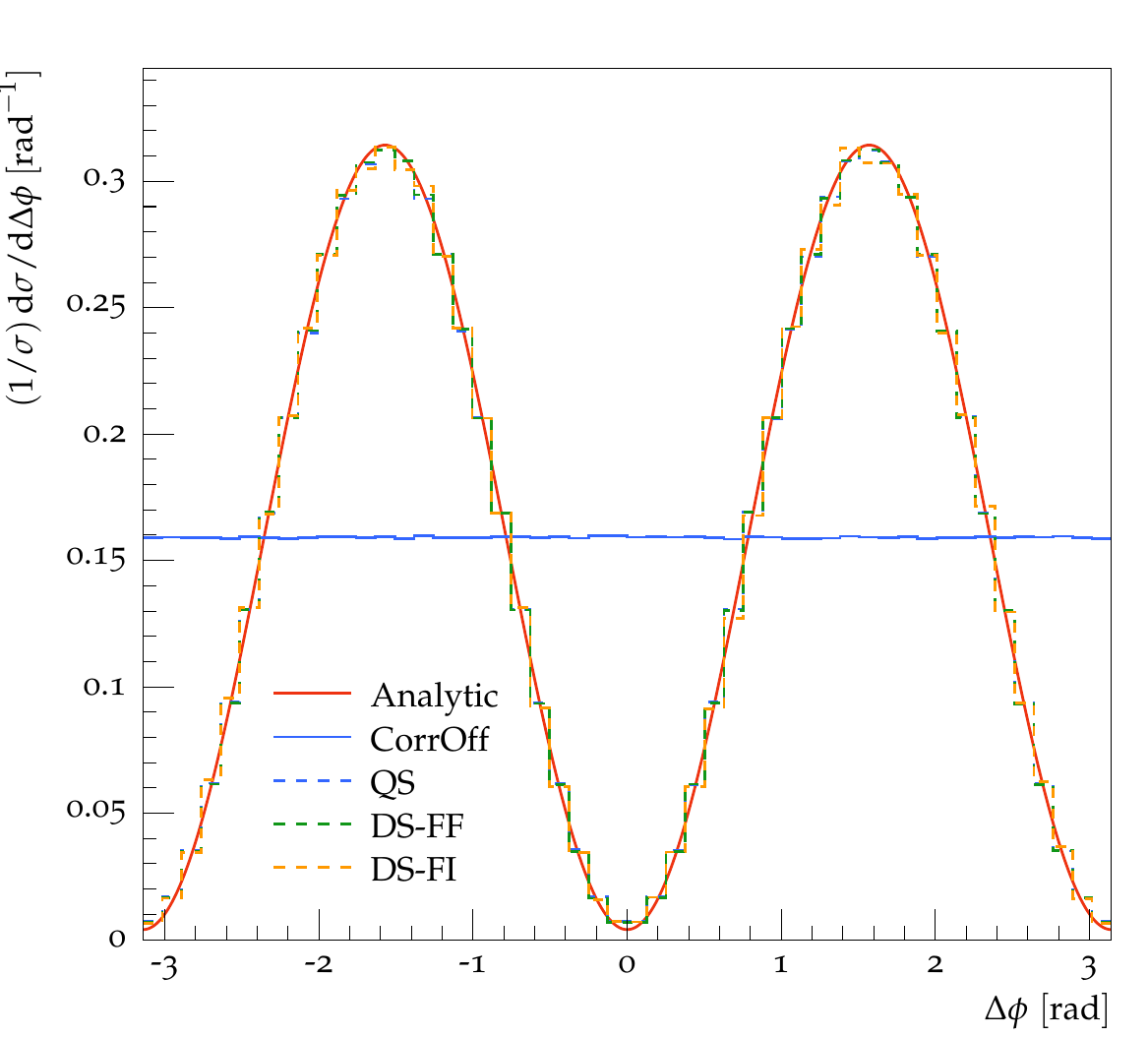}
  \end{center}
  \vspace{-10pt}
  \caption{The analytic result for the difference in azimuthal angle between
    the branching planes of subsequent final-state $g\to gg$ and $g\to q\bar{q}$ splittings
    compared to the distributions predicted using the angular-ordered~(QS) and
    dipole parton showers in \Herwig7. The predictions obtained using only final-final~(DS-FF)
    and final-initial~(DS-FI) dipoles in the dipole shower are shown separately.
    The prediction of the angular-ordered~(CorrOff) shower without spin correlations is included
    for comparison.
    The momentum fraction in the first and second branchings lies in the range
    $0.9 < z_1 < 1.0$ and $0.4 < z_2 < 0.5$ respectively.}
  \label{fig:FSR-NearestNeighbour-gg-qq}
\end{figure}

\begin{figure}
  \begin{center}
    \includegraphics[width=0.45\textwidth]{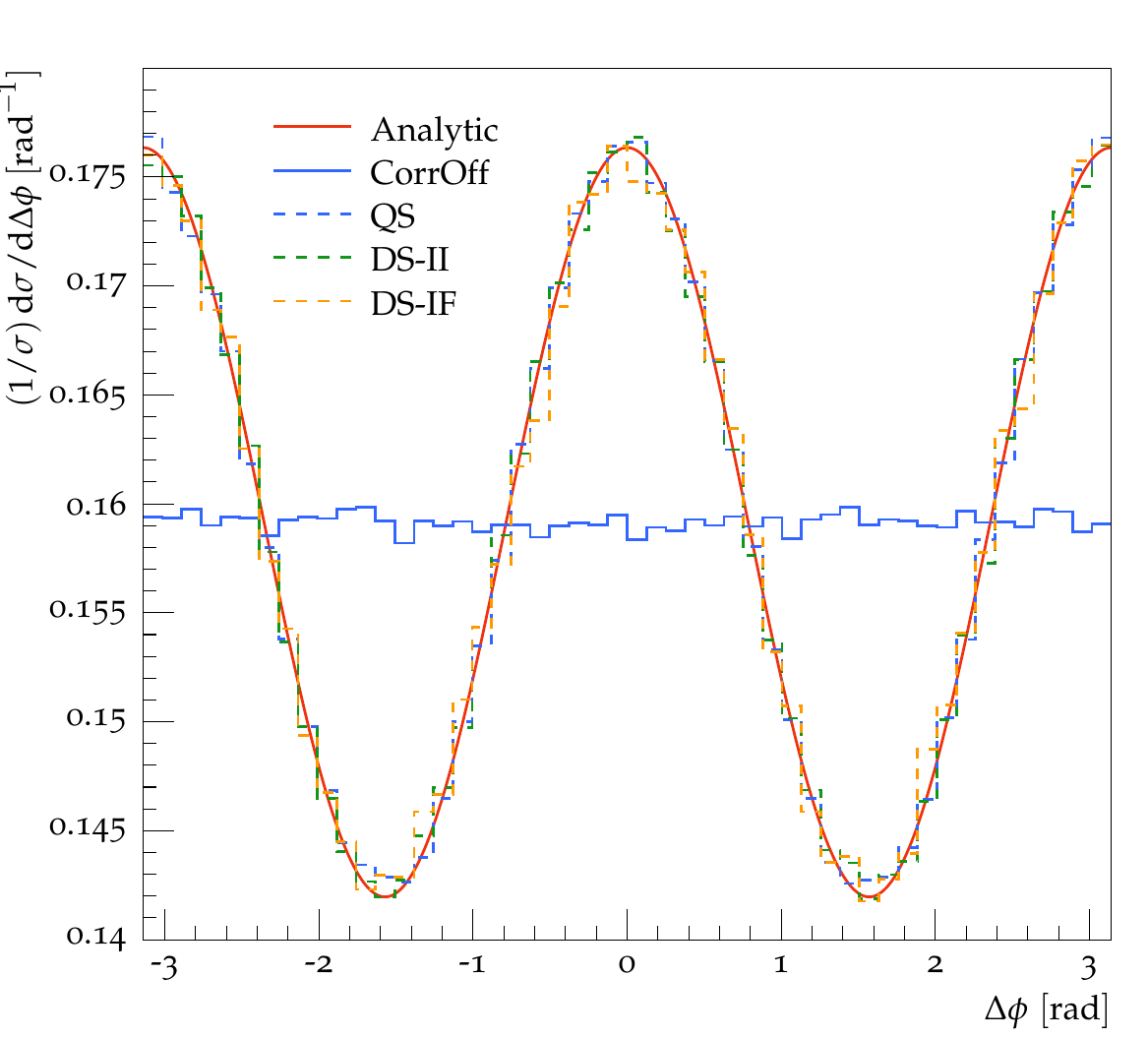}
  \end{center}
  \vspace{-10pt}
  \caption{The analytic result for the difference in azimuthal angle between
    the branching planes of subsequent initial-state $q\to qg$ and $g\to gg$ splittings
    compared to the distributions predicted using the angular-ordered~(QS) and
    dipole parton showers in \Herwig7. The predictions obtained using only initial-initial~(DS-II)
    and initial-final~(DS-IF) dipoles in the dipole shower are shown separately.
    The prediction of the angular-ordered~(CorrOff) shower without spin correlations is included
    for comparison.
    The momentum fraction in the first and second branchings lies in the range
    $0.9 < z_1 < 1.0$ and $0.4 < z_2 < 0.5$ respectively.}
  \label{fig:ISR-NearestNeighbour-qg-gg}
  \begin{center}
    \includegraphics[width=0.45\textwidth]{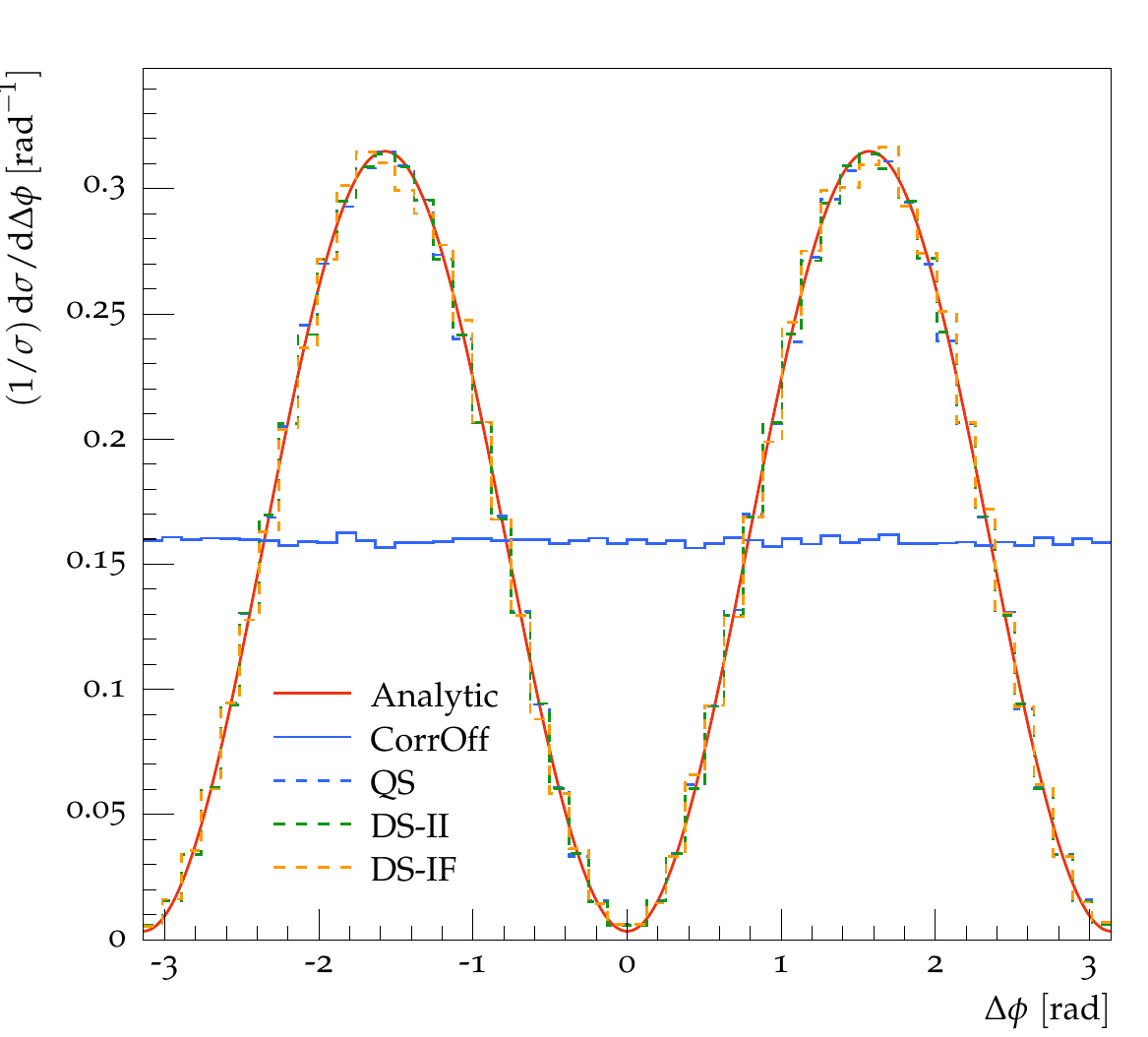}
  \end{center}
  \vspace{-10pt}
  \caption{The analytic result for the difference in azimuthal angle between
    the branching planes of subsequent initial-state $q\to qg$ and $g\to q\bar{q}$ splittings
    compared to the distributions predicted using the angular-ordered~(QS) and
    dipole parton showers in \Herwig7. The predictions obtained using only initial-initial~(DS-II)
    and initial-final~(DS-IF) dipoles in the dipole shower are shown separately.
    The prediction of the angular-ordered~(CorrOff) shower without spin correlations is included
    for comparison.
    The momentum fraction in the first and second branchings lies in the range
    $0.9 < z_1 < 1.0$ and $0.4 < z_2 < 0.5$ respectively.}
  \label{fig:ISR-NearestNeighbour-qg-qq}
\end{figure}

\begin{figure}
  \begin{center}
    \includegraphics[width=0.45\textwidth]{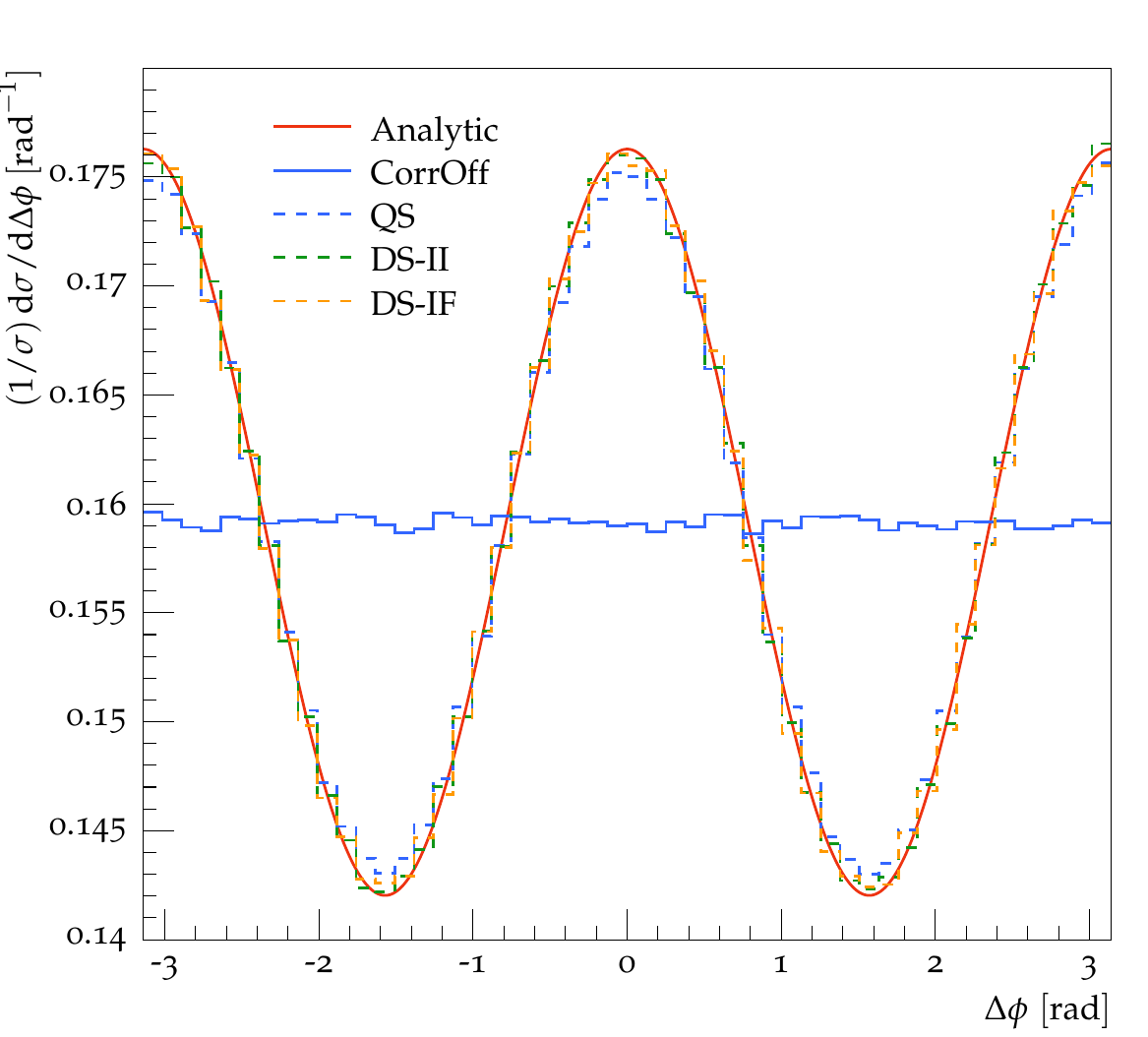}
  \end{center}
  \vspace{-10pt}
  \caption{The analytic result for the difference in azimuthal angle between
    the branching planes of subsequent initial-state $g\to gg$ and $g\to gg$ splittings
    compared to the distributions predicted using the angular-ordered~(QS) and
    dipole parton showers in \Herwig7. The predictions obtained using only initial-initial~(DS-II)
    and initial-final~(DS-IF) dipoles in the dipole shower are shown separately.
    The prediction of the angular-ordered~(CorrOff) shower without spin correlations is included
    for comparison.
    The momentum fraction in the first and second branchings lies in the range
    $0.9 < z_1 < 1.0$ and $0.4 < z_2 < 0.5$ respectively.}
  \label{fig:ISR-NearestNeighbour-gg-gg}
  \begin{center}
    \includegraphics[width=0.45\textwidth]{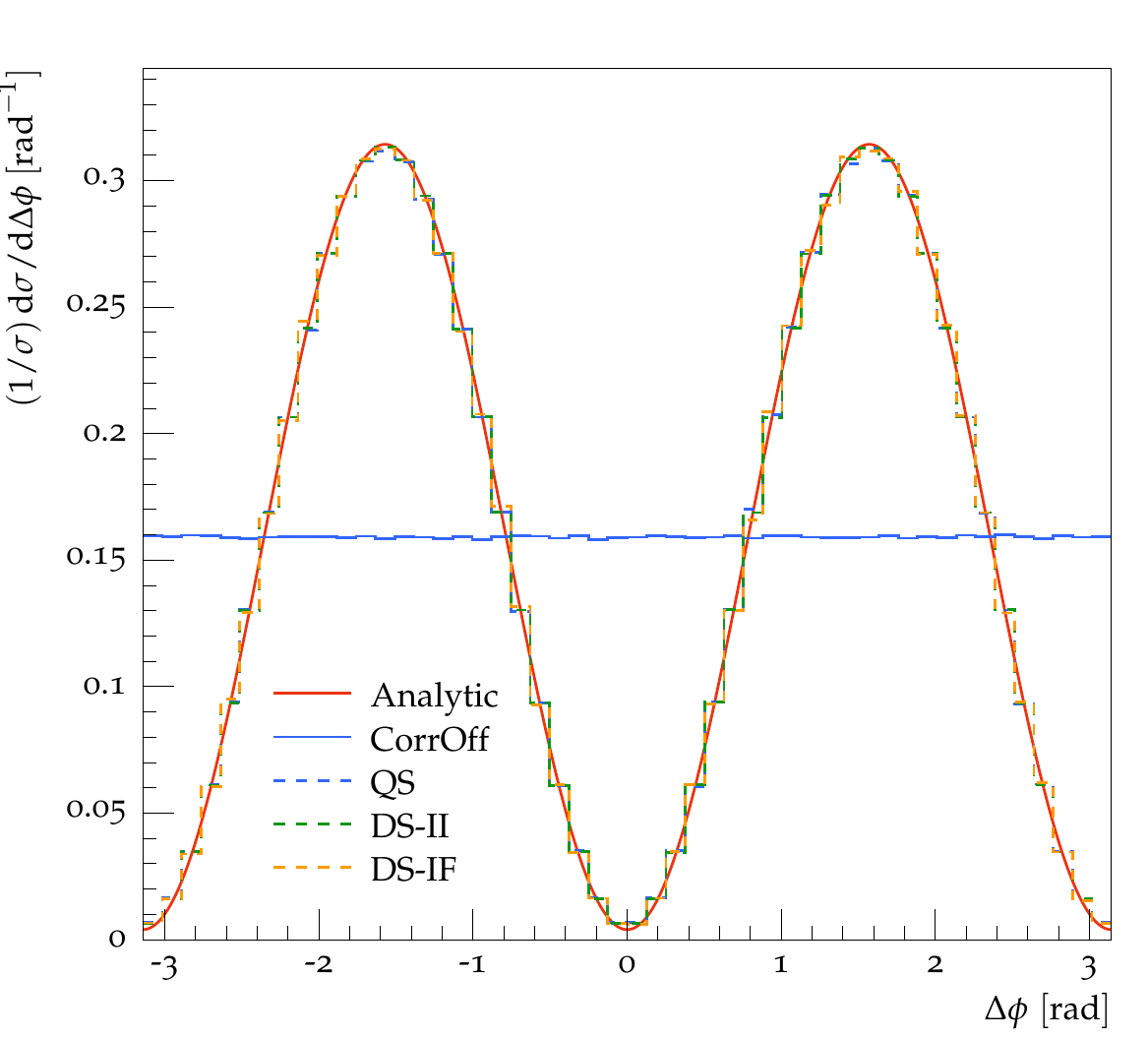}
  \end{center}
  \vspace{-10pt}
  \caption{The analytic result for the difference in azimuthal angle between
    the branching planes of subsequent initial-state $g\to gg$ and $g\to q\bar{q}$ splittings
    compared to the distributions predicted using the angular-ordered~(QS) and
    dipole parton showers in \Herwig7. The predictions obtained using only initial-initial~(DS-II)
    and initial-final~(DS-IF) dipoles in the dipole shower are shown separately.
    The prediction of the angular-ordered~(CorrOff) shower without spin correlations is included
    for comparison.
    The momentum fraction in the first and second branchings lies in the range
    $0.9 < z_1 < 1.0$ and $0.4 < z_2 < 0.5$ respectively.}
  \label{fig:ISR-NearestNeighbour-gg-qq}
\end{figure}

The results are shown in
Figs.\,\ref{fig:FSR-NearestNeighbour-qg-gg}-\ref{fig:FSR-NearestNeighbour-gg-qq} and
Figs.\,\ref{fig:ISR-NearestNeighbour-qg-gg}-\ref{fig:ISR-NearestNeighbour-gg-qq}
for FSR and ISR respectively.
We have chosen to measure the azimuthal-difference for splittings in which the momentum fraction
in the first and second branchings lies in the range $0.9 < z_1 < 1.0$ and $0.4 < z_2 < 0.5$
respectively as this is the configuration in which the correlation is strongest.
All of the results shown are for the case of massless quarks.

Each plot shows the analytic result and the parton shower predictions.
In each plot we have included the prediction obtained using the angular-ordered shower with
spin correlations switched off. In each case this gives rise to a flat line at $1/2\pi$ and
we have confirmed that the dipole shower also predicts a flat line.
All of the parton shower predictions display good agreement with the analytic
result in each case.


\subsection{Correlations with the Hard Process}
\label{sect:corr-in-hard-proc}

In this section we consider results that probe the correlations between the parton shower and the
hard process. These tests verify that correlations are passed correctly between the hard
process and the parton shower.
In addition these tests also probe the treatment of spectators and splitting recoils
in the dipole shower, as discussed in Section\,\ref{sect:dipole-shower}.

\begin{figure}
  \begin{center}
    \includegraphics[width=0.45\textwidth]{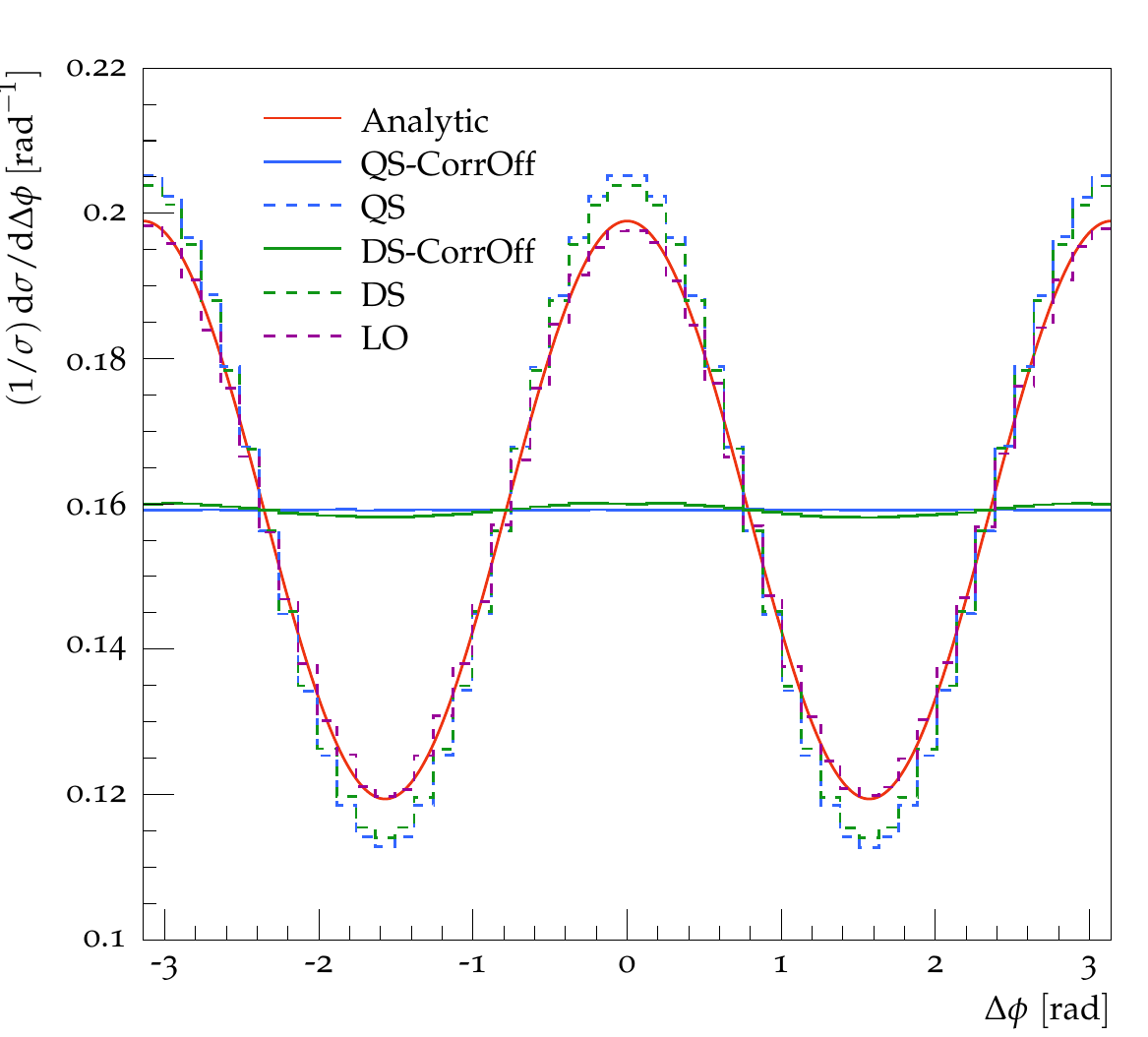}
  \end{center}
  \vspace{-10pt}
  \caption{The analytic result for the difference in azimuthal angle between the planes
    of the two branchings in \mbox{$h^0\to g g \to q \bar{q} q'\bar{q'}$} compared to the distributions
    predicted using the angular-ordered~(QS) and dipole~(DS) parton showers in \Herwig7.
    The angular-ordered shower~(QS-CorrOff) and dipole shower~(DS-CorrOff) predictions without spin correlations
    are included for comparison.
    The result obtained from a sample of LO events generated using \MG~(LO) is also shown. }
  \label{fig:FSR-dphi}
\end{figure}

The analytic result for the distribution of the azimuthal angle between the planes of the
$g\to q\bar{q}$ branchings in \mbox{$h^0\to g g \to q \bar{q} q'\bar{q'}$} is given in
Eqn.~\eqref{eq:h0-gg-qqqq}. This analytic result and the predictions of
 the angular-ordered and dipole parton showers are shown in Fig.\,\ref{fig:FSR-dphi}. In
addition we include the result obtained from a sample of leading-order (LO) events
generated using \MG~\cite{Alwall:2014hca}. All quarks are massless and our analysis requires
two gluon splittings to different quark flavours to enable perfect identification of the quark
pairs.
The shower predictions both display good agreement with the analytic result and the LO prediction.
The differences that remain are due to the cutoff on the transverse momentum used in both parton
showers, whereas the analytic result has no cutoff and the LO result includes a cut on the invariant
mass of the quark-antiquark pairs which does not affect the shape of the distribution. The transverse
momentum cutoff removes more of the region $z\to0,1$ where the correlation is smallest giving a slightly
larger correlation effect overall.

The above result probes the the treatment of FSR. In order to test the correlations between
ISR and the hard process we also consider the Higgs boson production process
$gg \to h^0$ followed by the backward splitting of each of the two gluons into an incoming quark
and an outgoing quark. In order to obtain a finite leading-order result we require that the minimum
transverse momentum of the outgoing quarks is 20\,GeV.

\begin{figure}
  \begin{center}
    \includegraphics[width=0.45\textwidth]{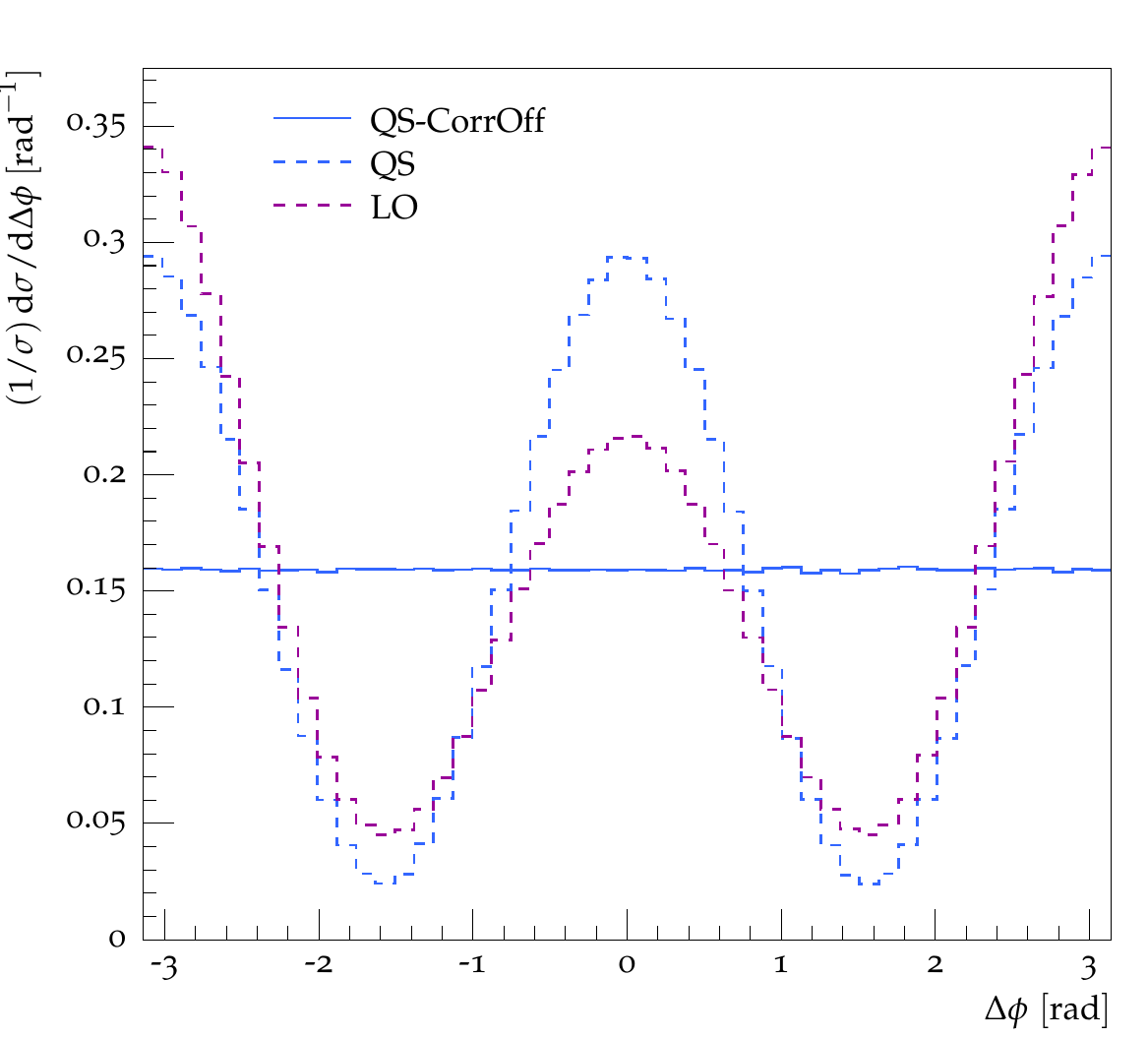}
  \end{center}
  \vspace{-10pt}
  \caption{The difference in azimuthal angle between the planes
    of two initial-state $g \to q\bar{q}$ branchings in $g g \to h^0$ predicted using
    the angular-ordered~(QS) parton shower in \Herwig7.
    The angular-ordered parton shower~(QS-CorrOff) prediction without spin correlations is also
    included.
    The result obtained from a sample of LO events generated using \MG~(LO) is shown
    for comparison. }
  \label{fig:ISR-dphi-QS}
\end{figure}

\begin{figure}
  \begin{center}
    \includegraphics[width=0.45\textwidth]{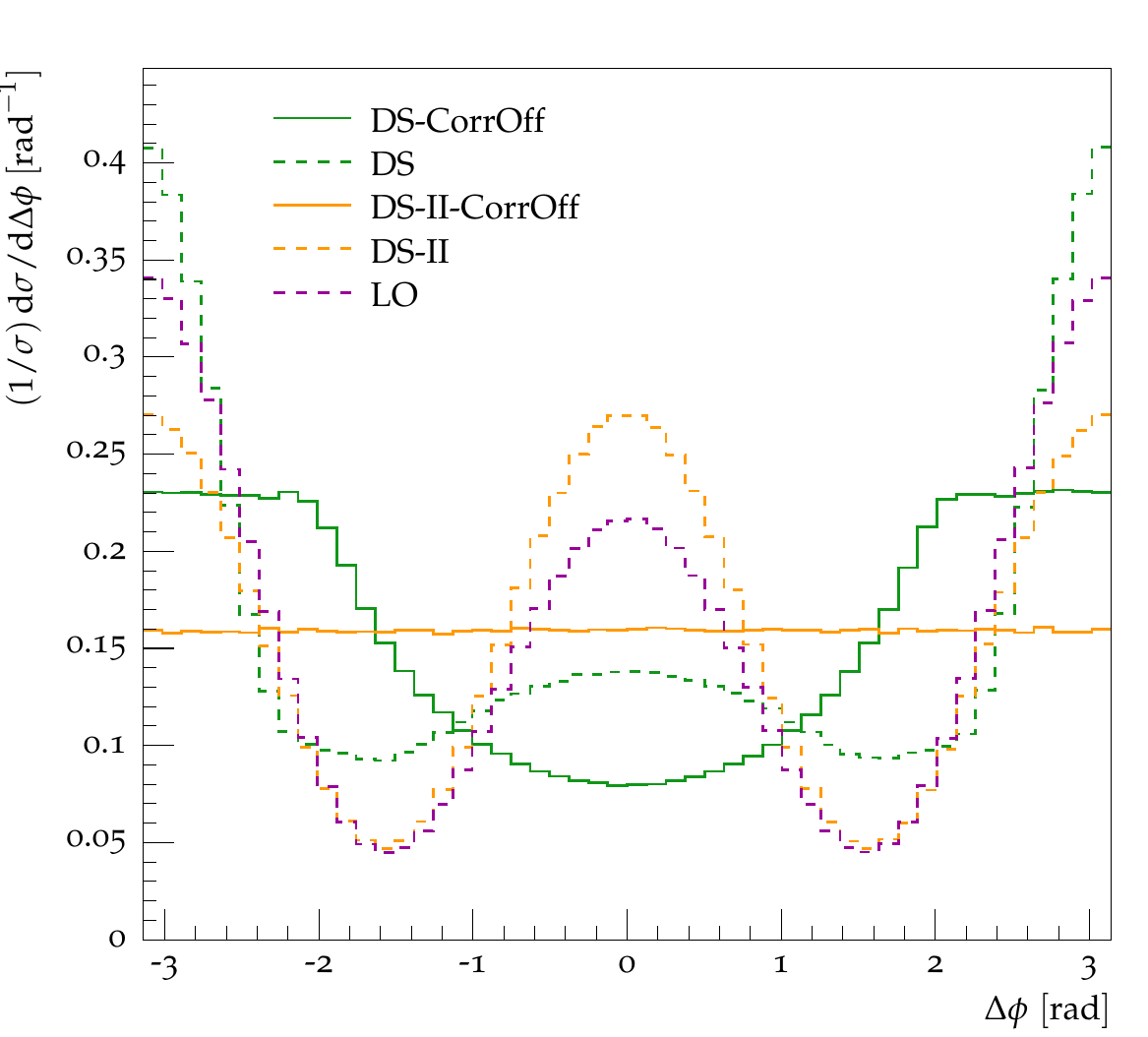}
  \end{center}
  \vspace{-10pt}
  \caption{The difference in azimuthal angle between the planes
    of two initial-state $g \to q\bar{q}$ branchings in \mbox{$g g \to h^0$} predicted using 
    the dipole parton shower~(DS) in \Herwig7. The dipole parton shower~(DS-CorrOff) prediction without
    spin correlations is also included.
    Predictions obtained using the dipole parton shower restricted to allow branchings from
    II dipoles only and with a modified handling of splitting recoils, as described in the text,
    are shown with~(DS-II) and without~(DS-II-CorrOff) spin correlations.
    The result obtained from a sample of LO events generated using \MG~(LO) is shown
    for comparison. }  
  \label{fig:ISR-dphi-DS}
\end{figure}

The predictions for the distribution of the difference in the azimuthal angle between
the planes of the branchings
predicted using the angular-ordered and dipole parton showers are shown in
Fig.\,\ref{fig:ISR-dphi-QS} and Fig.\,\ref{fig:ISR-dphi-DS} respectively.
In each plot we also include the result obtained from a sample of LO events
generated using \MG\ for comparison.

In Fig.\,\ref{fig:ISR-dphi-QS} we include predictions obtained using the angular-ordered parton shower
with and without spin correlations. When spin correlations are not included in the parton shower
the predicted distribution is simply a flat line. We find that, with spin correlations included
in the parton shower, the angular-ordered parton shower prediction is similar to the LO prediction with
some differences in shape due to corrections beyond the collinear limit.

The predictions obtained using the dipole parton\linebreak shower display more complex behaviour and
we have included several results in Fig.\,\ref{fig:ISR-dphi-DS}.
We first note that the prediction produced using the dipole parton shower without
spin correlations is not flat. This is due to the treatment of splitting recoils.
In initial-initial dipoles the recoil in splittings is distributed amongst all outgoing
particles other than the new emission, and in
initial-final dipoles the outgoing spectator gains a transverse component to its momentum.
The momentum of the outgoing quark produced in the first splitting is therefore changed in
a non-trivial way in the second splitting and this gives rise to a directional preference of
the second splitting relative to the first splitting. This behaviour necessarily affects the
prediction when spin-correlations are included and gives rise to the corresponding distribution in
Fig.\,\ref{fig:ISR-dphi-DS}.

In order to demonstrate that the effects seen in the dipole parton shower predictions are indeed due
to the treatment of recoil momenta in splittings, we have also included results obtained
using a modified version of the dipole parton shower. In this modified parton shower we only allow splittings
off initial-initial dipoles and we have modified the behaviour of these splittings such that
the splitting recoil is entirely absorbed by the outgoing Higgs boson in both of the splittings.
With these modifications the direction of the quark produced in the first splitting is not modified in the
second splitting and when spin correlations are not included the predicted distribution is a flat line.
As such the prediction with spin correlations included displays better
agreement with the angular-ordered parton shower and LO predictions. Again there are differences in shape
between the dipole parton shower prediction and the LO prediction due to corrections beyond the collinear limit.

Similar problems with the default recoil scheme in\linebreak dipole parton showers
were recently observed in Ref.\,\cite{Dasgupta:2018nvj} where it was shown
that the same change in the recoil strategy we have used resolved issues with the
logarithmic accuracy of the parton shower.

\subsection{Correlations in Decay Processes}
\label{sect:corr-in-decays}
\begin{figure}
  \begin{center}
    \includegraphics[width=0.45\textwidth]{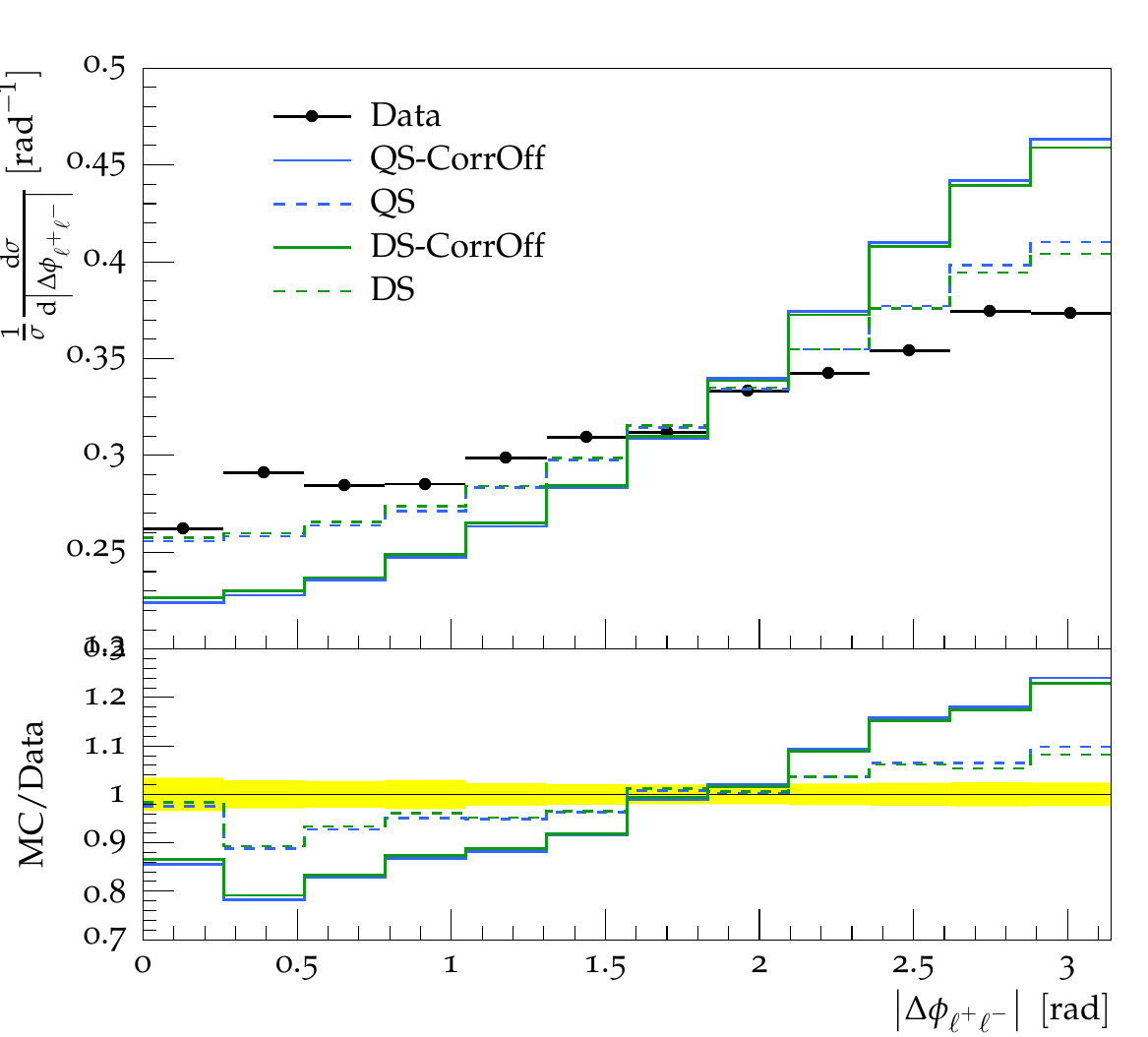}
  \end{center}
  \vspace{-10pt}
  \caption{The azimuthal separation of the charged leptons in 8 TeV dileptonic $pp\to t\bar{t}$
    events, as measured by CMS~\cite{Khachatryan:2016xws} and predicted using the
    angular-ordered~(QS) and dipole~(DS) parton showers in \Herwig7.
    The predictions of the angular-ordered~(QS-CorrOff) shower and the dipole shower~(DS-CorrOff) 
    without spin correlations are also shown.}
  \label{fig:pp-ttbar}
\end{figure}

The spin correlations in the hard process can also affect the distribution of the particles
produced in the subsequent decay of unstable particles, such as the top quark, and also
give correlations between the decay products of different unstable particles.
As we are interested in correlations in the parton shower,
in this section we look at correlations in the decay of a coloured particle, namely the top
quark. Fig.\,\ref{fig:pp-ttbar} shows the azimuthal separation of the charged leptons in
dileptonic $pp\to t\bar{t}$ events at a centre-of-collision energy of 8 TeV, measured by CMS.
In addition we show the predictions of this distribution obtained using the angular-ordered
and dipole parton showers in \Herwig7. The hard process is produced using a LO matrix element.
In the angular-ordered shower the top-quark decays are corrected to NLO in QCD while in the
dipole shower no such correction is applied to obtain these predictions.
There is reasonable agreement between the experimental result and both parton shower algorithms including
spin correlations whereas the results without spin correlations clearly fail to describe the data.

\section{Conclusions}

We have implemented the spin correlation algorithm of
Refs.\,\cite{Collins:1987cp,Knowles:1987cu,Knowles:1988hu,Knowles:1988vs,Richardson:2001df} in the
angular-ordered and dipole parton showers in \Herwig7. This feature will be available for public use in \Herwig7.2.
We have compared the predictions obtained using each of the parton showers in \Herwig7 to
analytic calculations or predictions obtained using a LO ME. Through these comparisons we
have confirmed that the spin correlation algorithm is functioning correctly in both showers.

The handling of splitting recoils in the dipole shower is not formally included in the spin
correlation algorithm. We have discussed these limitations and presented results that
show where these effects are evident.
Despite these limitations, we find that the dipole
shower, and the angular-ordered shower, produce a fairly accurate prediction of a
spin-correlation sensitive observable, namely the azimuthal separation of the leptons,
in $pp\to t\bar{t}$ events.

While spin correlation effects are often unobservable in average distributions, as we have seen
there are cases
where they are important. Their implementation in \Herwig7 is therefore an important part of improving
the accuracy of the simulation.

\section*{Acknowledgements}

We thank the other members of the Herwig collaboration, and Simon Pl\"atzer in particular,
for useful discussions and support.  SW acknowledges support from a STFC studentship.
This work received funding from the European
Union’s Horizon 2020 research and innovation programme as
part of the Marie Sk\l{}odowska-Curie Innovative Training Network MCnetITN3 (grant agreement no. 722104).

\label{sec:Conclusions}

\appendix

\section{Quasi-collinear Spin Unaveraged Splitting Functions}
\label{app:split}

The calculation of the spin unaveraged splitting functions for
$g\to q \bar{q}$ is presented in Section\,\ref{sect:branching}.
We present the calculation of the remaining spin unaveraged
splitting functions here.

\subsection{Branching $g\to g g$}

In this case $m=m_1=m_2=0$ and the polarization vectors for the outgoing gluons
are:
\begin{eqnarray}
  \epsilon^\mu_{\lambda_1} &=& -\frac{\lambda_1}{\sqrt{2}}  \left(0;
  1-{\frac {p_\perp^2\lambda^2 {\rm e}^{i\lambda_1\phi} \cos\phi }{2{{\bf p}}^2{z}^2}},
  \right.\\ &&\ \ \ \ \ \ \ \ \ \left.
  i\lambda_1-{\frac {{p_\perp}^2\lambda^2{\rm e}^{i\lambda_1\phi}\sin\phi}{2{\bf p}^2z^2}},
  -\frac{p_\perp{\rm e}^{i\lambda_1\phi}}{{\bf p}z}\lambda
  \right) +\mathcal{O}(\lambda^3);\nonumber\\
  \epsilon^\mu_{\lambda_2} &=& -\frac{\lambda_2}{\sqrt{2}}\left(0;
  1         -\frac {p_\perp^2\lambda^2{\rm e}^{ i\lambda_2\phi} \cos \phi  }{2{\bf p}^2 \left(1-z\right)^2},
  \right.\\ &&\ \ \ \ \ \ \ \ \ \left.
  i\lambda_2-\frac {p_\perp^2\lambda^2{\rm e}^{ i\lambda_2\phi} \sin \phi  }{2{\bf p}^2 \left(1-z\right)^2},
  \frac{p_\perp{\rm e}^{i\lambda_2\phi}}{\left(1-z \right) {\bf p}}\lambda
  \right) +\mathcal{O}(\lambda^3).\nonumber
\end{eqnarray}
The helicity amplitudes for the branching can be written as 
\begin{eqnarray}
   F_{\lambda_0\lambda_1\lambda_2} &=& 
\sqrt{\frac2{q^2_0}}
\left[
	 q_1\cdot\epsilon^*_2  \epsilon_0 \cdot \epsilon^*_1
	-q_1\cdot\epsilon_0  \epsilon^*_1 \cdot \epsilon^*_2
	-q_2\cdot\epsilon^*_1  \epsilon_0 \cdot \epsilon^*_2
\right],\nonumber\\
  &=& -\frac12\sqrt{z(1-z)}e^{i(\lambda_0-\lambda_1-\lambda_2)\phi'}\nonumber\\&&
\left[
	 \lambda_0\lambda_1\lambda_2\left(\frac1{z(1-z)}-1\right)
	+\lambda_0+\frac{\lambda_1}{z}+\frac{\lambda_2}{1-z}
\right].\nonumber\\
\end{eqnarray}
   This function gives the correct spin averaged splitting function and also
   is the same as the splitting functions used in \Herwig6 from \cite{Knowles:1987cu}~Table~1.
   The results for the individual helicity states are given in Table\,\ref{tab:gluon}.

\begin{table}
\begin{center}
\begin{tabular}{|c|c|c|c|c|}
\hline
$\lambda_0$ & $\lambda_1$ & $\lambda_2$ & $g\to g g$ & $g\to q \bar{q}$ \\
\hline
 + & + & + & $-{\frac {1}{\sqrt {z \left( 1-z \right) }}}$&
	${\frac {m}{z \left( 1-z \right) \tilde{q}}}$\\
\hline
 + & + & - & ${\frac {{z}^{3/2}}{\sqrt {1-z}}}$&
	$z\sqrt {1-{\frac {{m}^2}{{z}^2 \left( 1-z \right) ^2\tilde{q}^2}}}
$\\
\hline
 + & - & + & ${\frac { \left( 1-z \right) ^{3/2}}{\sqrt {z}}}$&
	$- \left( 1-z \right) \sqrt {1-{\frac {{m}^2}{{z}^2 \left( 1-z \right) ^2\tilde{q}^2}}}
$\\
\hline
 + & - & - & $0$&
	$0$\\
\hline
 - & + & + & $0$&
	$0$\\
\hline
 - & + & - & $-{\frac { \left( 1-z \right) ^{3/2}}{\sqrt {z}}}$&
	$\left( 1-z \right) \sqrt {1-{\frac {{m}^2}{{z}^2 \left( 1-z
 \right) ^2\tilde{q}^2}}}
$\\
\hline
 - & - & + & $-{\frac {{z}^{3/2}}{\sqrt {1-z}}}$&
	$-z\sqrt {1-{\frac {{m}^2}{{z}^2 \left( 1-z \right) ^2\tilde{q}^2}}}
$\\
\hline
 - & - & - & ${\frac {1}{\sqrt {z \left( 1-z \right) }}}$&
	${\frac {m}{z \left( 1-z \right) \tilde{q}}}$	\\
\hline
\end{tabular}
\caption{Spin unaveraged splitting functions for $g\to gg$ and $g\to q \bar{q}$.
         In addition to the factors given above each term has a phase
	$e^{i\phi'(\lambda_0-\lambda_1-\lambda_2)}$, where $\lambda_{1,2}=\pm1$ for outgoing gluons
 	and $\lambda_{1,2}=\pm\frac12$ for outgoing quarks. For the $g\to q\bar{q}$
	splitting function $\tilde{q}$ is the \Herwig7 evolution variable.}
\label{tab:gluon}
\end{center}
\end{table}

\subsection{Quark and Antiquark Branching}

  For quarks and antiquarks there is only one branching to consider, {\it i.e.} $q\to q g$.
  In this case $m_0=m_1=m$ and $m_2=0$.
  This gives the spinors:
\begin{subequations}
\begin{eqnarray}
u_{\frac12}(p) &=& \left(\begin{array}{c}
\frac {m}{\sqrt{2{\bf p}}}\lambda\\ 0 \\ \sqrt {2{\bf p}}\left(1+\frac{m^2\lambda^2}{8{\bf p}^2}\right) \\ 0
\end{array}\right);\\ 
u_{-\frac12}(p) &=& \left(\begin{array}{c}
  0\\ \sqrt {2{\bf p}}\left(1+\frac {m^2\lambda^2}{8{\bf p}^2}\right)\\ 0 \\\frac{m}{\sqrt{2{\bf p}}}\lambda
\end{array}\right).
\end{eqnarray}
\label{eqn:qqguin}
\end{subequations}
 
The polarization vectors of the gluon are
\begin{eqnarray}
  \epsilon^\mu_{\lambda_2}(q_2) &=& \left[0;
    -\frac{\lambda_2}{\sqrt{2}}\left(1-\frac {p_\perp^2 \lambda^2{\rm e}^{i\lambda_2\phi} \cos\phi}{2{\bf p}^2\left(1-z\right)^2}\right),\right.\nonumber\\&&\!\!\!\!\!\left.
    -\frac{i}{\sqrt{2}} +\frac {\lambda_2p_\perp^2\lambda^2{\rm e}^{i\phi} \sin\phi}{2\sqrt {2}{\bf p}^2 \left(1-z\right)^2},
   -\frac { \lambda_2p_\perp\lambda{\rm e}^{i\lambda_2\phi} }{\sqrt {2} \left(1-z\right) {\bf p}}\right]\!\!.
 \label{eqn:qqgeps}
\end{eqnarray}
and the spinors of the outgoing quark:
\begin{subequations}
  \begin{eqnarray}
    \bar{u}_{\frac12}(q_1)  &=&
    \left[\sqrt {2z{\bf p}}\left(1+\frac{m^2\lambda^2}{8{\bf p}^2}\right),
      \frac {{\rm e}^{-i\phi} p_\perp\lambda}{\sqrt {2z{\bf p}}},
      \right.\nonumber\\&&\ \ \ \ \ \left.
      \frac {m}{\sqrt {2z{\bf p}}}\lambda,
      \frac {{\rm e}^{-i\phi}p_\perp m{\lambda}^2}{2z{\bf p}\sqrt{2z{\bf p}}}
      \right]\!; \\
    \bar{u}_{-\frac12}(q_1)  &=& \left[
      -\frac { {\rm e}^{i\phi} p_\perp m\lambda^2}{2z{\bf p}\sqrt{2z{\bf p}}},
      \frac{m}{\sqrt {2z{\bf p}}}\lambda,
      \right.\nonumber\\&&\ \ \ \ \ \left.
-\frac { {\rm e}^{i\phi} p_\perp\lambda}{\sqrt{2z{\bf p}}},
\sqrt {2z{\bf p}}\left(1+\frac {m^2\lambda^2}{8z{\bf p}}\right)
\right]\!.
  \end{eqnarray}
  \label{eqn:qqgout}
\end{subequations}

In this case the helicity amplitudes for the branching are
\begin{equation}
  F_{\lambda_0\lambda_1\lambda_2} = \sqrt{\frac1{2(q^2_0-m^2)}}\bar{u}_{\lambda_1}(q_1)\epsilon_{\lambda_2}\slac 
	u_{\lambda_0}(q_0).
\end{equation}
  These functions are given in Table~\ref{tab:quark}.\footnote{If we make the same redefinition as before these functions agree with those in \cite{Knowles:1987cu}~Table~1 up to
the overall sign which is irrelevant.}

\begin{table}
  \begin{center}
    \begin{tabular}{|c|c|c|c|}
      \hline
      & & \multicolumn{2}{c|}{$F_{\lambda_0,\lambda_1,\lambda_2}$} \\
      \cline{3-4}
      $\lambda_0$ & $\lambda_1$ & $\lambda_2=+$ &$\lambda_2=-$  \\
      \hline
      + & + & ${\frac {1}{\sqrt {1-z}}\sqrt {1-{\frac {{m}^2}{{z}^2{\tilde{q}}^2}}}}$ &
              $-{\frac {z}{\sqrt {1-z}}\sqrt {1-{\frac {{m}^2}{{z}^2{\tilde{q}}^2}}}}$\\
      \hline
      + & - & $\frac{m}{\tilde{q}}\frac{\sqrt{1-z}}{z}$ &
              $0$\\
      \hline
      - & + & $0$ &
              $\frac{m}{\tilde{q}}\frac{\sqrt{1-z}}{z}$ \\
      \hline
      - & - & ${\frac {z}{\sqrt {1-z}}\sqrt {1-{\frac {{m}^2}{{z}^2{\tilde{q}}^2}}}}$ &
              $-{\frac {1}{\sqrt {1-z}}\sqrt {1-{\frac {{m}^2}{{z}^2{\tilde{q}}^2}}}}$\\
      \hline
    \end{tabular}
  \end{center}
\caption{Spin unaveraged splitting functions for $q\to qg$.
         In addition to the factors given above each term has a phase
	$e^{i\phi'(\lambda_0-\lambda_1-\lambda_2)}$, where $\lambda_{2}=\pm1$ for the outgoing gluon
 	and $\lambda_{0,1}=\pm\frac12$ for quarks and 
	 $\tilde{q}$ is the \Herwig++ evolution variable.}
\label{tab:quark}
\end{table}
\subsection{Squark Branching}

For squarks there is only one branching to consider, {\it i.e.} $\tilde{q}\to\tilde{q}g$.
The kinematics are the same as those for gluon radiation from a quark, {\it i.e.}
$m_0=m_1=m$ and $m_2=0$.
The polarization vectors of the gluons are also the same as for radiation from a quark,
Eqn.\,\ref{eqn:qqgeps}.

The spin unaveraged splitting function is given by
\begin{eqnarray}
 F_{\lambda_2} &=& \sqrt{\frac2{q^2_0-m^2}} q_1\cdot \epsilon_2^* \nonumber\\
            &=& \lambda_2 e^{-i\lambda_2\phi'}
	\sqrt{1-\frac{m^2}{z^2\tilde{q}^2}}\sqrt{\frac{z}{1-z}}.
\end{eqnarray}
If we sum over the helicities of the gluon we obtain the correct 
massive splitting function.

\section{Basis Rotation}
\label{sect:rotation}

In the case that we have a spin density matrix from the production process and then wish to generate the
correlations in the shower for simplicity we will have to deal with the rotation of the basis
states used.

Consider a process with production matrix element $\mathcal{A}_{\rm prod}$ and a matrix element $\mathcal{A}_{\rm branch}$ giving the 
subsequent time-like branching of the intermediate particle with momentum $p$ and mass $m$.

In the case of a fermion we will replace the spin sum with the Dirac spin sum, {\it i.e}
\begin{equation}
   p\sla + m  = \sum_{a}\Psi_a\bar\Psi_a,
\end{equation}
or similarly for vector bosons
\begin{equation}
  -g^{\mu\nu}+\frac{p^\mu n^\nu+p^\nu n^\mu}{p\cdot n} = \sum_a\epsilon_a^\mu\epsilon_a^{*\nu}
\end{equation}

We write both of these as
\begin{equation}
  \sum_a\chi_a\chi_a^{\dagger},
\end{equation}
where the basis state for the production is $\chi^\dagger_a$ and for the branching is 
$\chi_a$ such that the matrix element for the full process is
\begin{equation}
\mathcal{M} = \mathcal{A}_{\rm branch} \chi_a\chi_a^\dagger\mathcal{A}_{\rm prod},
\end{equation}
using the Einstein convention.

The matrix element squared is then
\begin{equation}
  |\overline{\mathcal{M}}|^2 =
  \chi^\dagger_b\mathcal{A}^\dagger_{\rm branch}\mathcal{A}_{\rm branch} \chi_a
  \chi_a^\dagger\mathcal{A}_{\rm prod}\mathcal{A}^\dagger_{\rm prod}\chi_b
\end{equation}
If we normalise by the spin summed matrix element for the production process then we obtain
\begin{equation}
  \frac{|\overline{\mathcal{M}|^2}}{\overline{|\mathcal{M}_{\rm prod}|^2}} =  \rho_{ab}\chi^\dagger_b\mathcal{A}^\dagger_{\rm branch}\mathcal{A}_{\rm branch} \chi_a,
\end{equation}
where $\mathcal{M}_{\rm prod}=\chi_a^\dagger\mathcal{A}_{\rm prod}$ and
\begin{equation}
  \rho_{ab} = \frac1{\overline{|\mathcal{M}_{\rm prod}|^2}}\chi_a^\dagger\mathcal{A}_{\rm prod}\mathcal{A}^\dagger_{\rm prod}\chi_b.
\end{equation}

In order to simplify the calculation we have made a specific choice of the basis of the
wavefunctions $\chi^\prime_a$ when we calculate the branching, which may not be the same as those used
in the calculation of the production, $\chi_a$, after Lorentz transformation into the frame in which the branching is calculated.

We therefore need to calculate the rotation between these two choices of basis states
\begin{equation}
\chi_a = \omega_{ai}\chi^\prime_i,
\end{equation}
therefore the calculation of the distribution is given by
\begin{eqnarray}
  \frac{|\overline{\mathcal{M}|^2}}{\overline{|\mathcal{M}_{\rm prod}|^2}} &=&    \omega_{ai} \rho_{ab}\omega^*_{bj}\chi^{\prime\dagger}_j\mathcal{A}^\dagger_{\rm branch}\mathcal{A}_{\rm branch} \chi^\prime_i \\
  &=& \rho^\prime_{ij} \chi^{\prime\dagger}_j  \mathcal{A}^\dagger_{\rm branch}\mathcal{A}_{\rm branch} \chi^\prime_i,
  = \rho^\prime_{ij} \mathcal{M}^\dagger_{{\rm branch} j} \mathcal{M}_{{\rm branch} i}\nonumber
\end{eqnarray}
where
\begin{equation}
 \rho^\prime_{ij} = \omega_{ai} \rho_{ab}\omega^*_{bj} 
\end{equation}
and $\mathcal{M}_{{\rm branch} i} = \mathcal{A}_{\rm branch} \chi^\prime_i$.

If we consider the case of space-like branching then the incoming particle to the hard process will now have basis state $\chi_a$ and
outgoing particle from the space-like branching $\chi^\dagger_a$ such that the matrix element is
\begin{equation}
\mathcal{M} = \mathcal{A}_{\rm prod} \chi_a\chi_a^\dagger\mathcal{A}_{\rm branch},
\end{equation}

The matrix element squared is then
\begin{equation}
  |\overline{\mathcal{M}}|^2 =
  \chi^\dagger_b\mathcal{A}^\dagger_{\rm prod}\mathcal{A}_{\rm prod} \chi_a
  \chi_a^\dagger\mathcal{A}_{\rm branch}\mathcal{A}^\dagger_{\rm branch}\chi_b
\end{equation}
If we normalise by the spin summed matrix element for the production process then we obtain
\begin{equation}
  \frac{|\overline{\mathcal{M}|^2}}{\overline{|\mathcal{M}_{\rm branch}|^2}} =  D_{ab}\chi^\dagger_a\mathcal{A}_{\rm branch}\mathcal{A}^\dagger_{\rm branch} \chi_b,
\end{equation}
where in this case $\mathcal{M}_{\rm prod}=\mathcal{A}_{\rm prod}\chi_a$ and
\begin{equation}
  D_{ab} = \frac1{\overline{|\mathcal{M}_{\rm prod}|^2}} \chi^\dagger_b\mathcal{A}^\dagger_{\rm prod}\mathcal{A}_{\rm prod} \chi_a.
\end{equation}
If we then rotate the basis states as before
the calculation of the distribution is given by
\begin{eqnarray}
  \frac{|\overline{\mathcal{M}|^2}}{\overline{|\mathcal{M}_{\rm prod}|^2}} &=&
  \omega_{bj}D_{ab}\omega^*_{ai}\chi^{\prime^\dagger}_i\mathcal{A}_{\rm branch}\mathcal{A}^\dagger_{\rm branch} \chi^\prime_j
  \\
  &=& 
  D^\prime_{ij}\chi^{\prime^\dagger}_i\mathcal{A}_{\rm branch}\mathcal{A}^\dagger_{\rm branch} \chi^\prime_j = 
  D^\prime_{ij}\mathcal{M}_{{\rm branch} i} \mathcal{M}^\dagger_{{\rm branch} j}\nonumber
\end{eqnarray}
where
\begin{equation}
 D^\prime_{ij} =\omega_{bj}D_{ab}\omega^*_{ai} 
\end{equation}
and $\mathcal{M}_{{\rm branch} i} = \chi^{\prime^\dagger}_i\mathcal{A}_{\rm branch}$.

\bibliography{spin}

\end{document}